\documentclass[12pt, draftclsnofoot, onecolumn]{IEEEtran}

\usepackage{subfloat}
\usepackage{setspace}
\usepackage{amsmath}
\usepackage{amsfonts}
\usepackage{graphicx}
\usepackage{amsthm}
\usepackage{amssymb}
\usepackage{mathrsfs}
\usepackage{stfloats}
\usepackage{color}
\usepackage{bm}
\usepackage{subfigure}
\usepackage[caption=false,font=footnotesize]{subfig}
\usepackage{cite}
\usepackage{cases}

\newtheorem{Lemma}{Lemma}
\newtheorem{Theorem}{Theorem}

\newtheorem{Proposition}{Proposition}
\theoremstyle{definition}
\newtheorem{Observation}{Observation}
\newtheorem{Remark}{Remark}

\begin{document}
        \title{On Performance of Distributed RIS-aided Communication in Random Networks}
		
\author{Jindan~Xu,~\IEEEmembership{Member,~IEEE}, Wei~Xu,~\IEEEmembership{ Senior~Member,~IEEE},\\and Chau Yuen,~\IEEEmembership{ Fellow,~IEEE}			 
\thanks{
}


\thanks{J. Xu and C. Yuen are with the School of Electrical and Electronics Engineering, Nanyang Technological University, Singapore
639798, Singapore (e-mail: jindan.xu@ntu.edu.sg, chau.yuen@ntu.edu.sg).}
\thanks{W. Xu is with the National Mobile Communications Research Laboratory, Southeast University, Nanjing 210096, China, and also with Purple Mountain Laboratories, Nanjing 211111, China (e-mail: wxu@seu.edu.cn). }

%
}

\maketitle
	
\begin{abstract}
This paper evaluates the geometrically averaged performance of a wireless communication network assisted by a multitude of distributed reconfigurable intelligent surfaces (RISs), where the RIS locations are randomly dropped obeying a homogeneous Poisson point process.
By exploiting stochastic geometry and then averaging over the random locations of RISs as well as the serving user, we first derive a closed-form expression for the spatially ergodic rate in the presence of phase errors at the RISs in practice.
Armed with this closed-form characterization, we then optimize the RIS deployment under a reasonable and fair constraint of a total number of RIS elements per unit area. The optimal configurations in terms of key network parameters, including the RIS deployment density and the array sizes of RISs, are disclosed for the spatially ergodic rate maximization.
Our findings suggest that deploying larger-size RISs with reduced deployment density is theoretically preferred to support extended RIS coverages, under the cases of bounded phase shift errors.
However, when dealing with random phase shifts, the reflecting elements are recommended to spread out as much as possible, disregarding the deployment cost.
\color{black}
Furthermore, the spatially ergodic rate loss due to the phase shift errors is quantitatively characterized.
For bounded phase shift errors, the rate loss is eventually upper bounded by a constant as $N\rightarrow\infty$, where $N$ is the number of reflecting elements at each RIS.
While for random phase shifts, this rate loss scales up in the order of $\log N$.
These analytical observations are validated through numerical results.
\end{abstract}
	

	\IEEEpeerreviewmaketitle

\section{Introduction}
%
%
%
%

Reconfigurable intelligent surface (RIS) is recently arising as a promising technology to meet numerous requirements of emerging wireless applications in the sixth-generation (6G) networks, e.g., Internet-of-Everything \cite{WShi_IoE}, integrated sensing and communications, and semantic communication \cite{WXu_Semantic}.
It is commonly easy to integrate an RIS into wireless environment to establish favorable reflecting links, e.g., via RIS deployment into an indoor ceiling, a building facade, and an unmanned aerial vehicle~\cite{ XCao_AT}.
Comprising of a vast amount of cost-efficient reflecting elements, RIS is capable of dynamically manipulating the propagation environment by adjusting the phase shift of each reflecting element in real time via a smart controller \cite{MDRenzo_JSAC}. Alternatively, extra information can also be carried via phase shift modulation methods by RIS~\cite{JYao_Modu, MWen_Modu, JYao_Modu_TWC}.
Serving as either a reflector or a transmitter, RIS has shown great potentials in improving spectral efficiency, energy efficiency, coverage, and security \cite{CHuang_Holo,SCIS,WShi_TWC}.

Typically, RIS is acknowledged with the ability of constructing supplementary links from the transmitter towards the desired receiver to strengthen the received signal power \cite{CPan_Overview, WTang}.
These supplementary links play an important role in assisting wireless communications especially when the direct line-of-sight (LoS) path between the transceivers is blocked by obstacles such as buildings, trees, and human bodies \cite{Stat_CSI_Coverage_Yang}.
In order to fully reap the benefits of RIS, intensive efforts have been invested to the optimization of phase shift design to maximize multifarious performance metrics under various constraints \cite{MWen_OFDM, BDi_Rates_Disc, CHuang_EE, QWu_JointBeamf, CHuang_DRL}.
In \cite{BDi_Rates_Disc}, an iterative algorithm was proposed to maximize the sum rate of an RIS-aided multiuser downlink channel by jointly optimizing the RIS phase shifts, transmit precoding, and power allocation.
In \cite{CHuang_EE}, energy efficiency was maximized by the joint optimization of RIS phase shifts and transmit precoding under the constraints of Quality-of-Service requirements.
Similarly in \cite{QWu_JointBeamf}, the joint optimization was formulated to target transmit power minimization subject to the constraint of received signal-to-interference-and-noise ratio (SINR).
Unlike the alternating convex optimization methods used in these works, a deep reinforcement learning (DRL)-based algorithm was proposed in \cite{CHuang_DRL} to simultaneously acquire both the optimal phase shifts and transmit precoding as the outputs of a neural network.

From another perspective of RIS performance evaluation, there have been increasing attention with intensive studies \cite{QWu_JointBeamf, Stat_CSI_Coverage, Stat_BoardCoverage}.
To be specific, it was revealed in \cite{QWu_JointBeamf} that the RIS-enhanced signal power scales up in the order of $N^2$ where $N$ is the number of RIS reflecting elements.
Further in \cite{Stat_CSI_Coverage}, it was evinced that the coverage probability scales up with $N^2$ and $N$ under the condition of the optimal and random phase shifts, respectively.
Besides, the downlink rate of a millimeter-wave broadcast channel was proven to increase logarithmically with the power gain reflected by RIS, which is exploited to combat severe attenuation of high-frequency transmission \cite{Stat_BoardCoverage}.


The aforementioned studies mostly rely on the assumption of perfect channel state information (CSI) at the transmitter.
However, since RISs are passive devices, they cannot actively transmit or receive pilot signals, making direct channel estimation for RIS channels infeasible.
To address this issue, a number of effective estimation methods have been developed to acquire the RIS-assisted channels \cite{CE_CYou, CE_Nadeem, CE_ZWang, CE_ZQHe, CE_Araújo, CE_Taha}.
For instance, a straightforward framework for the cascade channel estimation, called ON/OFF method, was developed in \cite{CE_CYou, CE_Nadeem, CE_ZWang}, which splits the entire reflecting array into multiple sub-arrays before estimating each sub-channel separately.
In \cite{CE_ZQHe, CE_Araújo}, matrix decomposition technique was applied for estimating sparse RIS multiple-input multiple-output (MIMO) channels.
Recently, deep learning algorithms were also exploited to facilitate RIS channel estimation \cite{CE_Taha}.
However, these methods require demanding overhead and excessive hardware cost, which is usually hard to achieve in practice especially for large RISs \cite{LWei_CE,WZhang_CE,LWei_CE_TCOM}.
Thus, it is still challenging to acquire perfect CSI with affordable cost in large RIS-assisted MIMO communications.
Recently, there have been initial efforts devoted to robust design and performance evaluation for RIS-aided communication systems with imperfect CSI \cite{WZhang_CSI, JYao_CSI, Stat_CSI__MultiPair, Stat_CSI_2Timescale} and practical hardware constraints \cite{DYang_RIS_Hardware, PhaseError, JXu_Disc, CHuang_Disc, yhan, LYou_Tradeoff, SZhou_SE_EE_Impair}.
In \cite{Stat_CSI__MultiPair}, an RIS-aided multi-pair communication system was investigated with the knowledge of statistical CSI, which is slow-varying and much easier to track compared to instantaneous CSI.
Alternatively, a two-timescale transmission scheme was proposed in \cite{Stat_CSI_2Timescale}, where the RIS phase shifts were designed based on statistical CSI while the transmit percoding was designed according to instantaneous CSI.
In \cite{CHuang_Disc}, low-resolution phase shifts were considered in a multiuser multiple-input single-output (MISO) network. It was discovered that the RIS equipped with even 1-bit phase shifters outperforms a conventional amplify-and-forward relay since the power consumption of a passive RIS is negligible.
Moreover in \cite{yhan}, the required phase shift resolution for a tolerable spectral efficiency loss was analyzed for an RIS-aided MISO system.
Considering both statistical CIS and discrete phase shifts, the study \cite{LYou_Tradeoff} characterized the tradeoff between energy efficiency and spectral efficiency in an RIS-aided multiuser MIMO system.
In addition, the impact of hardware impairments on an RIS-aided downlink network was studied in \cite{SZhou_SE_EE_Impair}. Due to the radio-frequency impairments, it was revealed that the spectral efficiency eventually saturates even if the number of RIS elements grows to infinity.



Recently, multi-RIS-aided communication has garnered increasing attention due to its potential benefits \cite{YZhao}.
A study in \cite{yuweigao} demonstrated that a distributed multi-RIS deployment outperforms a centralized single-RIS architecture under the constraint of the same number of total reflecting elements.
This superiority arises because distributed RISs offer additional reflecting paths between the transceivers compared to a single centralized RIS.
Besides, the phase shifts of distributed RISs were optimized to maximize the sum rate and SINR in \cite{Dist_DRL} and \cite{Dist_Aghashahi}, respectively.
Moreover, the authors in \cite{Dist_Do} proposed two purpose-oriented multi-RIS-aided schemes, called the exhaustive RIS-aided and opportunistic RIS-aided approaches.
In particular, the exhaustive approach utilizes all available RISs to assist communication, while the opportunistic approach selectively employs the most appropriate RIS.
In addition to the aforementioned considerations of fixed RIS positions, stochastic geometry was utilized to model random drops of RISs \cite{beam5g, jlyu, suijiris1, exploitrandom, Dist_Shafique}.
By modeling the random locations of RISs by a Binomial Point Process (BPP), the authors in \cite{Dist_Shafique} successfully analyzed the coverage probability, energy efficiency, and ergodic capacity of the network.
It was revealed that a higher density is beneficial to the communication at the expense of higher hardware cost, while the tradeoff between the performance and cost is still unknown.
In \cite{jlyu}, the locations of RISs were modeled as a homogeneous Poisson point process (HPPP) using a boolean model.
By averaging over these random locations, the spatially ergodic achievable rate of an RIS-aided downlink network was characterized.
Then in \cite{suijiris1}, the locations of both RISs and base stations (BSs) were modeled by independent HPPPs. 
It was revealed that the energy efficiency first degrades and then grows if either the deployment density of RIS or the density of BS increases.
In addition to the locations of RISs and BSs, those of blockages were modeled by an HPPP in \cite{exploitrandom}, while the length, width, and orientation of each blockage were drawn from uniform distributions.
Under this assumption, the probability of the presence of a reflecting path generated by an RIS was successfully derived.
However, most of these existing studies assumed distributed RISs with ideal phase shifts and perfect CSI, which is hard to achieve in practice.
The obtained expressions of spatially ergodic spectral efficiency are usually intractable due to the involved integrals with the distributions of random locations.


This paper investigates a downlink network assisted with a multitude of distributed RISs.
In particular, we theoretically analyze the spatially ergodic rate of the network averaged over random locations of the RISs and user equipment (UE).
In addition, since it is difficult to implement ideally continuous phase shifts at the RIS due to the imperfect CSI and the employment of discrete phase shifters in practice, we characterize the impact of the phase shift errors on the spatially ergodic rate.
Main contributions of this paper are summarized as follows.

$\bullet$
By applying the technique of stochastic geometry and then averaging over the random locations of RISs and UE,
we derive the spatially ergodic rate in closed form for a distributed RIS-aided network, where the locations of distributed RISs are stochastically modeled by an HPPP.
To the best of our knowledge, this paper conducts a first attempt to evaluate the geometrically averaged performance of a multi-RIS-aided network while incorporating the impact of imperfect phase shifts at RISs.


$\bullet$
Due to the constraint of hardware cost in practice, we introduce a fair and tractable constraint of a total number of RIS elements per unit area, that is $\lambda N \equiv \eta$ for a constant $\eta$.
Under this constraint, it is crucial to strike a balance between the RIS size, $N$, and the RIS deployment density, $\lambda$, for spatially ergodic rate maximization.
By maximizing the ergodic rate, we derive the optimal values of $N$ and $\lambda$ in closed forms, which scale like $N^*=\frac{\eta}{\lambda^*}\sim \mu^{-1} C^2$ where $C$ is the serving radius of RIS and $\mu \triangleq \frac{\sin\left(\rho\pi\right)}{4\rho}$ depends on $\rho\in[0,1)$ as the level of phase shift errors.
Under the condition of bounded phase shift errors, it implies that the reflecting elements are better used to form larger RISs with reduced density for a larger serving radius $C$.
While for random phase shifts, the optimal values are obtained in closed form as $N^*=\frac{\eta}{\lambda^*}=1$, regardless of $C$.

$\bullet$
We characterize the spatially ergodic rate loss caused by imperfect phase shifts.
Particularly for the worst-case scenario of $\rho=1$, corresponding to the uniformly random phase shift design, the rate loss scales up logarithmically with respect to (w.r.t.) $N$.
However, for phase shift errors within a bounded range with $\rho\in[0,1)$,
the rate loss eventually saturates to a constant of $2\left(1-\mathrm{e}^{-\pi\lambda C^2}\right)\log \frac{\pi}{4\mu}$ as $N\rightarrow\infty$.
Taking the 1-bit coarse phase shifters for example, the corresponding rate loss is upper bounded by 1.3~bps/Hz with a typical $\rho=0.5$.

The remainder of this paper is organized as follows.
The system model of a downlink network assisted with distributed RISs is described in Section~\ref{Sec_Sys}.
In Section~\ref{Sec_ErgR}, we analyze the ergodic rate of a cell-edge UE given fixed RIS and UE locations.
Using stochastic geometry in Section~\ref{Sec_Spa_ErgR}, we further characterize the spatially ergodic rate which is averaged over the random locations of RISs and UE.
Section~\ref{Sec_Sim} presents numerical results before concluding remarks in Section~\ref{Sec_Con}.

\textit{Notations:}	
$a\sim\mathcal{CN}(0,\sigma^2)$ represents a complex Gaussian variable with zero mean and covariance $\sigma^2$.
$\mathcal{U}[a,b]$ denotes a uniform distribution between $a$ and $b$, and
$\mathcal{R}(\sigma)$ is a Rayleigh distribution with scale parameter $\sigma$.
Operator $\angle \cdot$ represents the angle of a complex-value variable and $\lceil\cdot\rceil$ is the ceil function.
$\max\{a,b\}$ and $\min\{a,b\}$ returns the maximum and minimum of $a$ and $b$, respectively.
$\mathbb{E}\{\cdot\}$ and $\mathbb{V}\{\cdot\}$ denote the statistical expectation and variance operators, respectively.
$\mathcal{\gamma}(\alpha,x) \triangleq \int_0^x \mathrm{e}^{-t}t^{\alpha-1} ~\mathrm{d}t$ is the incomplete gamma function.

\section{System Model}
\label{Sec_Sys}

\begin{figure}
	\centering
	\includegraphics[width=.68\linewidth]{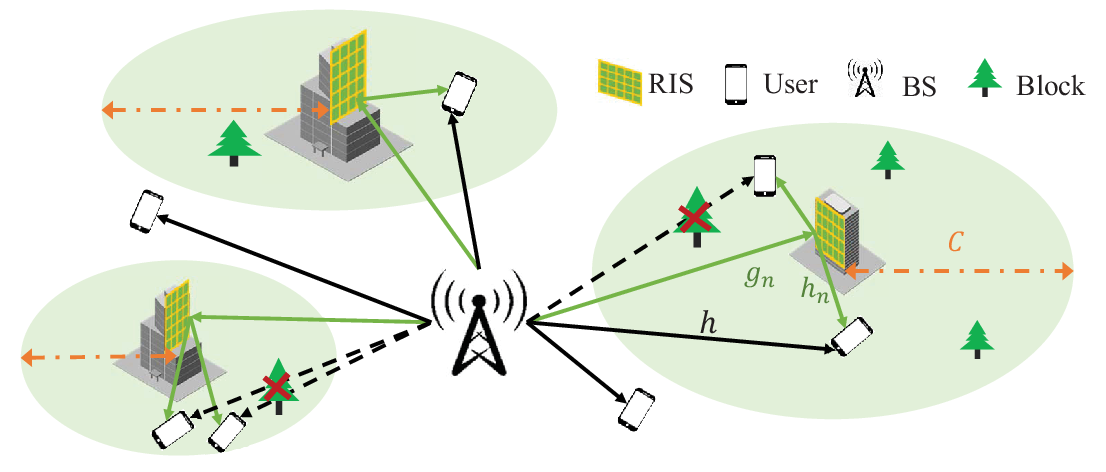}
	\caption{System model of a downlink network assisted with distributed RISs.}
	\label{fig:system}
\end{figure}

We consider a downlink network deploying a multitude of RISs to enhance the spectral efficiency, as illustrated in Fig.~\ref{fig:system}.
Assuming that UEs are uniformly located within the cell edge, we denote the distance between UE and the serving BS by $d\in[D_1, D_2]$, where $D_1$ and $D_2$ respectively represent the minimum and maximum distances.
The inter-cell interference is assumed ignorable by allocating orthogonal frequency bands to adjacent cell edges via common strategies like fractional frequency reuse and soft frequency reuse in LTE \cite{FFR, SFR}.
The positions of the RISs are modeled by an HPPP with density $\lambda$ \cite{exploitrandom},
where $\lambda$ is the average number of RISs per unit area.
Each RIS is a two-dimensional uniform planar array (UPA) integrated with $N=N_x N_y$ reflecting antenna elements, organized into $N_x$ rows and $N_y$ columns.
A typical association strategy is applied so that the UE is simultaneously served by its nearest RIS if it falls within the RIS serving area, i.e., $r\leq C$, where $C$ is defined as the RIS serving radius and $r$ is the distance between the UE and the nearest RIS.
Otherwise, if $r>C$, the UE is solely associated with the BS without RIS assistance.
The serving radius $C$ of RISs is assumed to be identical.
Considering the fact of a passive RIS being unable to transmit signals to identify the coverage area, the value of $C$ can be determined with the assistance of BS.
In addition, we consider that in current wireless networks the two dimensions of bandwidth and time within a given duration are divided into resource blocks (RBs), which are then allocated to distinct UEs. A single UE is served over a set of orthogonal RBs with other UEs \cite{jlyu}.
In cases where multiple UEs are within the serving area of the same RIS, they are assigned RBs at distinct time intervals. This temporal separation enables the RIS to serve a large number of UEs sequentially \cite{Dist_Shafique}.
Then, the received signal at the UE is expressed as
\begin{numcases}
    {y=} {}
    \sqrt{ P}\left(\beta_l^\frac{1}{2} \beta_r^\frac{1}{2}\sum_{n=1}^N g_{n}\mathrm{e}^{j\theta_{n}}h_{n}+\beta_d^\frac{1}{2} h\!\right)x
	\!+\!w
	\triangleq y_1,~~~~r\leq C,
    \label{Eq_y1_v0}\\
    \sqrt{ P}\beta_d^\frac{1}{2} hx+w
    \triangleq y_2,~~~~~~~~~~~~~~~~~~~~~~~~~~~~~~~~~~r> C,
	\label{Eq_y2}
\end{numcases}
where $\theta_{n}$ is the reflecting phase shift of the $n$th RIS element,
$h$ denotes the channel coefficient of the BS-UE link,
$g_{n}$ and $h_{n}$ respectively denote the BS-RIS and RIS-UE channel coefficients associated with the $n$th RIS element,
$\beta_l$, $\beta_r$, and $\beta_d$ respectively denote the large-scale fading coefficients of the BS-RIS, RIS-UE, and BS-UE channels,
$x$ is the transmit symbol with zero mean and unit variance,  $w\sim \mathcal{CN}(0,\sigma^2)$ denotes the thermal noise with energy $\sigma^2$, and $P$ is the transmit power.

When the phase of the reflecting channel through each RIS element and
the phase of the direct BS-UE channel, i.e., $\angle(g_{n}h_{n})$ and $\angle h$, are perfectly known at the BS, it is intuitively reasonable to align the phase shift by choosing $\theta_{n}=-\angle(g_{n}h_{n})+\angle h$ so that the reflecting and direct channels are co-phased to achieve constructive addition of the received signals from multiple paths at the UE~\cite{jlyu}.
However, it is usually challenging to acquire idea phase shift control in practice due the two reasons.
On one hand, perfect channel estimation and feedback imposes demanding requests on overhead and hardware cost, despite the availability of effective channel estimation methods \cite{LWei_CE,WZhang_CE,LWei_CE_TCOM}.
On the other hand, continuous phase shift control is hardly achievable because discrete phase shifters are commonly used at RISs, leading to quantized phase shift errors.
Under this condition, performance loss is dominated by the impact of finite-resolution phase shift quantization, especially for 1-bit phase shifters, compared to that of CSI imperfection at a reasonably-high signal-to-noise ratio (SNR).
Therefore, it is reasonable to assume that the phase shift error follows a distribution similar to the discrete quantization error \cite{CHuang_Disc, JXu_Disc, BDi_Rates_Disc}.
Assuming that popular uniformly-discrete phase shifters are exploited and the channels are stochastically isotropic, the quantization error follows a uniform distribution \cite{Disc_QWu}.
That is, the phase shift error at the $n$th RIS element, denoted by $\tau_{n}$, follows an independent and identical uniform distribution, i.e., $\tau_{n}\sim\mathcal{U}(-\rho\pi,\rho\pi)$, where $\rho\in[0,1]$ represents the range of phase error.
Note that for discrete phase shifters, we have $\rho=2^{-b}$ where $b$ is the number of quantization bits. In particular, a special choice of $\rho=1$ corresponds to the typical case of uniformly random phase shifts.
By substituting $\theta_{n}=-\angle(g_{n}h_{n})+\angle h+\tau_{n}$ into \eqref{Eq_y1_v0}, the received signal $y_1$ equals
\begin{equation}		
	y_1=\sqrt{ P}\left(\beta_l^\frac{1}{2} \beta_r^\frac{1}{2}\sum_{n=1}^N |g_{n}||h_{n}| \mathrm{e}^{j\tau_{n}}+\beta_d^\frac{1}{2} |h|\right)x \mathrm{e}^{j\angle h}
	+w
	.
	\label{Eq_y1}
\end{equation}

To facilitate the analysis of ergodic rate in the next section, we assume that $h$, $g_{n}$, and $h_{n}$ are independent and identically distributed (i.i.d.) complex Gaussian variables following $\mathcal{CN}(0,1)$.
Then, $|h|$, $|g_{n}|$, and $|h_{n}|$ are i.i.d. Rayleigh variables with scale parameter  $\sqrt{1/2} $, i.e., $|h
|\sim\mathcal{R}(\sqrt{1/2})$, $|g_{n}|\sim\mathcal{R}(\sqrt{1/2})$, and $|h_{n}|\sim\mathcal{R}(\sqrt {1/2})$ \cite{jlyu}.
Besides, the large-scale fading coefficients depend on the distances between the transceivers, yielding
\begin{align}
    \beta_d=\beta d^{-\alpha_1}, ~ \beta_l=\beta l^{-\alpha_2}, ~\text{and} ~ \beta_r=\beta r^{-\alpha_3},
    \label{Eq_beta}
\end{align}
where $\beta$ denotes the large-scale fading coefficient at a reference distance,
$\alpha_1$, $\alpha_2$, and $\alpha_3$ respectively denote the pathloss exponents of the BS-UE, BS-RIS, and RIS-UE channels,
and $l$ is the distance between the BS and RIS.


\section{Ergodic Rate Analysis Given Fixed RIS Locations}
\label{Sec_ErgR}
Before characterizing the spatially ergodic rate averaged over the random locations of RISs and UE, we initiate by evaluating the achievable rate given fixed locations in this section, in the presence of phase shift errors.
This preliminary analysis is conducted first to facilitate quantitative characterization of the compensation relationships among essential system parameters, including $N$, $P$, and $\rho$, which is intractable after integrating over the distributions of random locations.

According to \eqref{Eq_y2} and \eqref{Eq_y1}, the ergodic achievable rate of the cell-edge UE is expressed as
\begin{numcases}
{R=} {}
    \mathbb{E}\left\{\log\left(1+ \frac{P}{\sigma^2}\left|\beta_l^\frac{1}{2} \beta_r^\frac{1}{2}\sum_{n=1}^N |g_{n}||h_{n}| \mathrm{e}^{j\tau_{n}}+\beta_d^\frac{1}{2} |h|\right|^2\right)\right\}
	\triangleq R_1,~~r\leq C,
    \label{Eq_R1}\\
    \mathbb{E}\left\{\log\left(1+ \frac{P}{\sigma^2}\beta_d |h|^2\right)\right\}
    \triangleq R_2,~~~~~~~~~~~~~~~~~~~~~~~~~~~~~~~~~~~r> C.
	\label{Eq_R2}
\end{numcases}
Given the distances among the UE, the nearest RIS, and the serving BS, the large-scale fadings $\beta_d$, $\beta_l$, and $\beta_r$ are temporary constants. Under this condition, the ergodic rates in \eqref{Eq_R1} and \eqref{Eq_R2} are averaged over the distribution of phase shift error $\tau_n$ and the distributions of small-scale fadings $h_n$, $g_n$, and $h$.
The ergodic rates are characterized with tight bounds in Theorem~\ref{Th_R_up}.

\begin{Theorem}
\label{Th_R_up}
Under the assumption of phase shift uncertainty $\tau_{n}\sim\mathcal{U}(-\rho\pi,\rho\pi)$,
the ergodic rates in \eqref{Eq_R1} and \eqref{Eq_R2} are bounded by 
\begin{numcases}
{R\leq} {}
    \!\log\left(\!1+ \frac{P}{\sigma^2}\left[\beta_l \beta_r \mu^2 N^2+\beta_l \beta_r\left(1\!-\!\mu^2\right)N
	+\beta_l^\frac{1}{2}\beta_r^\frac{1}{2}\beta_d^\frac{1}{2}\sqrt{\pi}\mu N
	+\beta_d\right]\right)
	\triangleq \bar{R}_1, ~r\leq C,
	\label{Eq_R1_up}\\
	\!\log\left(\!1+ \frac{P}{\sigma^2}\beta_d \right)
	\triangleq \bar{R}_2,~~~~~~~~~~~~~~~~~~~~~~~~~~~~~~~~~~~~~~~~~~~~~~~~~~~~~~~~~~~~~~~r> C,
    \label{Eq_R2_up}
\end{numcases}
where we define
\begin{align}
    \mu \triangleq \frac{\sin\left(\rho\pi\right)}{4\rho},~~\rho\in[0,1].
    \label{Eq_mu}
\end{align}

\end{Theorem}

\begin{IEEEproof}
See Appendix~\ref{Proof_Th1}.
\end{IEEEproof}

From \eqref{Eq_R1_up}, the asymptotic ergodic rate with large $N$ is further characterized in the following proposition.

\begin{Proposition}
For a large RIS, the ergodic rate in \eqref{Eq_R1_up} asymptotically tends to
\begin{align}
	\bar{R}_{1} &\rightarrow
	\log\left(1+ \beta_l \beta_r \mu^2 N^2\frac{P}{\sigma^2}\right)
	\\&\triangleq
	\log\left(1+ \beta_l \beta_r \xi\right)
	\label{Eq_R1_p3}
	,
\end{align}
where we define 
\begin{align}
    \xi\triangleq\mu^2 N^2\frac{P}{\sigma^2},
\end{align}
which represents an equivalent SNR at the UE.
\end{Proposition}
\begin{IEEEproof}
    After some basic manipulations, the ergodic rate in \eqref{Eq_R1_up} is rewritten as
    \begin{align}
	&\bar{R}_1=
        \log\!\left(\!\!1\!+\!\! \frac{P}{\sigma^2}\beta_l \beta_r \mu^2 N^2
	\!\!\left[\!1\!\!+\!\!\frac{1-\mu^2}{\mu^2N}
	\!\!+\!\!\sqrt{\frac{\pi\beta_d}{\beta_l\beta_r}}\frac{1}{\mu N}
	\!\!+\!\!\frac{\beta_d}{\beta_l \beta_r \mu^2 N^2}\!\right]\!\right)
	\!.
    \label{Eq_R1_prf}
    \end{align}
    Then, the desired result in \eqref{Eq_R1_p3} is obtained from \eqref{Eq_R1_prf} by utilizing the facts that $\frac{1-\mu^2}{\mu^2N}\rightarrow 0$, $\sqrt{\frac{\pi\beta_d}{\beta_l\beta_r}}\frac{1}{\mu N}\rightarrow 0$, and $\frac{\beta_d}{\beta_l \beta_r \mu^2 N^2}\rightarrow 0$, which come from the consideration of a large RIS with $N\gg \mathrm{max}\left\{\!\frac{1}{\mu^2}\!-\!1, \frac{1}{\mu}\sqrt{\frac{\pi\beta_d}{\beta_l\beta_r}}\right\}$.
\end{IEEEproof}

\begin{Remark}
\label{Re_R1}
    Form \eqref{Eq_R1_p3}, $\bar{R}_{1}$ remains invariant if $\xi$ keeps as a constant.
    It implies that \emph{a decreasing $\mu$ due to the phase shift error can be compensated by either increasing $N$ or enlarging SNR, i.e., $\frac{P}{\sigma^2}$, to ensure a constant of $\mu^2 N^2\frac{P}{\sigma^2}$.}
    In particular, if the ideal phase shifters are replaced by low-cost discrete ones with  2-bit (or 1-bit) phase quantization, it yields from \eqref{Eq_mu} that $\mu^2$ decreases from $\frac{\pi^2}{16}$ to $\frac{1}{2}$ (or $\frac{1}{4}$). Under this condition, $\bar{R}_{1}$ remains invariant either if $N$ increases by 1.1 times (or 1.6 times) or if SNR is lifted by additional 0.9~dB (or 3.9~dB).
\end{Remark}

In order to analyze the impact of phase shift error on the ergodic rate in \eqref{Eq_R1_up}, we first consider two typical cases with $\rho=0$ and $\rho=1$.
Under the assumption of perfect CSI and ideal phase shifts at the RIS, i.e., $\mu=\frac{\sin\left(\rho\pi\right)}{4\rho}\big|_{\rho=0}=\frac{\pi}{4}$, the ergodic rate in \eqref{Eq_R1_up} serves as an ideal benchmark expressed as
\begin{align}
	&\bar{R}_1^\mathrm{Ideal}=
        \log\!\left(\!\!1\!\!+\!\!\frac{P}{\sigma^2}\!\!\left[\!\beta_l \beta_r \frac{\pi^2}{16} N^2\!\!+\!\!\beta_l \beta_r\!\!\left(\!\!1\!\!-\!\!\frac{\pi^2}{16}\!\right)\!\!N
	\!\!+\!\!\beta_l^\frac{1}{2}\beta_r^\frac{1}{2}\beta_d^\frac{1}{2}\frac{\pi^\frac{3}{2}}{4} N
	\!\!+\!\!\beta_d\!\right]\!\right)
	\label{Eq_R1_up_perfect}
	\!\!.
\end{align}
Assuming that the RIS randomly selects phase shifts without the knowledge of CSI, i.e., $\mu=\frac{\sin\left(\rho\pi\right)}{4\rho}\big|_{\rho=1}=0$, the ergodic rate in \eqref{Eq_R1_up} serves as a lower-bound benchmark as
\begin{align}
	\bar{R}_1^\mathrm{Random}
	=\log\left(1+ \frac{P}{\sigma^2}\left[\beta_l \beta_r N+\beta_d\right]\right)
	\label{Eq_R1_up_random}
	.
\end{align}
Comparing \eqref{Eq_R1_up_perfect} with \eqref{Eq_R1_up_random}, the equivalent SNR at the cell-edge UE scales up with $N^2$ and $N$ under the conditions of ideal and random phase shifts, respectively.


\section{Spatially Ergodic Rate Analysis Using Stochastic Geometry}
\label{Sec_Spa_ErgR}
This section characterizes the spatially ergodic rate using stochastic geometry.
To maximize the derived spatially ergodic rate, the parameters for the deployment of distributed RISs, including the density and the size of each RIS, are further optimized.
In addition, the spatially ergodic rate loss caused by the RIS phase shift errors is analyzed.

\subsection{Spatially Ergodic Rate Averaged over Random Locations}
We consider the spatially ergodic rate defined as the ergodic rate in Theorem~\ref{Th_R_up} averaged over the random locations of the distributed RISs and UE.
From Theorem~\ref{Th_R_up}, the ergodic rate is derived as $\bar{R}_1$ in \eqref{Eq_R1_up} when the UE is located within the serving radius of its nearest RIS. Otherwise, the ergodic rate is expressed as $\bar{R}_2$ in \eqref{Eq_R2_up}.
Accordingly, we can express the spatially ergodic rate as
\begin{align}
    \Tilde{R} = \mathbb{E}_{l,d}\left\{ \int_{0}^{C} \bar{R}_1 f_r(r) ~\mathrm{d}r +\int_{C}^{\infty} \bar{R}_2 f_r(r) ~\mathrm{d}r  \right\}
    \label{Eq_Er_Exp}
    ,
\end{align}
where $f_r(r)$ is the probability density function (PDF) of $r$.
Since the UE is located at the cell edge near a serving RIS, it is reasonable to consider that the relationships of $d\gg r$ and $l=d$ hold tightly.
Assuming that the positions of the RISs are modeled by an HPPP with density $\lambda$ and the UE is uniformly located
within the cell edge, the spatially ergodic rate defined in \eqref{Eq_Er_Exp} is given in Theorem~\ref{Th_Erg_R_int}.

\begin{Theorem}
\label{Th_Erg_R_int}
The spatially ergodic rate defined in \eqref{Eq_Er_Exp} is expressed as
\begin{align}
\Tilde{R}=&
    \left(1-\mathrm{e}^{-\pi\lambda C^2}\right)
	\left[\log\left(\frac{P}{\sigma^2}\beta^2N\left(\mu^2 N+1\!-\!\mu^2\right) \right)
	-\frac{\alpha_2}{\ln 2} \left(\frac{D_2^2\ln D_2 -D_1^2\ln D_1 }{D_2^2-D_1^2} -\frac{1}{2}\right)\right]
 \nonumber\\&
	+\mathcal{G}(N,\mu)
    + \Tilde{R}_{2}
	-\frac{\alpha_3}{2\ln 2}
    \left[\mathrm{Ei} \left(-\pi\lambda C^2\right)
    -\mathrm{e}^{-\pi\lambda C^2} \ln C^2
    -\ln(\pi\lambda)-E_0\right],
    \label{Eq_Erg_R_int}
\end{align}
where $\mathrm{Ei}(x)\triangleq\int_{-\infty}^x \frac{\mathrm{e}^t}{t}~\mathrm{d}t$ is the exponential integral function, $E_0$ is the Euler's constant,
$\mathcal{G}(N,\mu)$ is defined as
\begin{align}
    &\mathcal{G}(N,\mu)\!\! \triangleq \!\!\!
    \int_{D_1}^{D_2}\!\!\!\!\!\int_{0}^{C}\!\!\!\log\!\!\left(\!\!1
	\!\!+ \!\!\frac{\sqrt{\pi\beta d^{\alpha_2\!-\!\alpha_1}r^{\alpha_3} }\mu N
	\!+\! d^{\alpha_2\!-\!\alpha_1} r^{\alpha_3} \!+ \!(\sigma^2/P\beta) d^{\alpha_2} r^{\alpha_3} }
	{\beta\mu^2 N^2+\beta\left(1-\mu^2\right)N} \!\!\right) \!\!2\pi\lambda r \mathrm{e}^{-\pi\lambda r^2}\!\! \frac{2d}{D_2^2\!-\!D_1^2}~\mathrm{d}r\mathrm{d}d
    ,\label{Eq_G_int}
\end{align}
and
\begin{align}
    \Tilde{R}_2 \!
    \!=\!\!\int_{D_1}^{D_2} \!\!\int_{C}^{\infty}\!\!\log\left(\!1\!+\! \frac{P}{\sigma^2} \beta d^{-\alpha_1} \!\right) \! 2\pi\lambda r \mathrm{e}^{-\pi\lambda r^2} \frac{2d}{D_2^2\!-\!D_1^2}~\mathrm{d}r\mathrm{d}d
    .
    \label{Eq_R2_Er_int}
\end{align}
\end{Theorem}
\begin{IEEEproof}
See Appendix~\ref{Proof_Th_R_int} with preliminaries in Appendix~\ref{Ap_Useful}.
\end{IEEEproof}

Notice that it is challenging to directly get insightful observations from the derived spatially ergodic rate in \eqref{Eq_Erg_R_int} due to the integrals involved in terms $\mathcal{G}(N,\mu)$ and $\Tilde{R}_2$.
Therefore, we seek to deriving the closed-form expressions of $\Tilde{R}$ for high and low SNRs, respectively, given in Theorem~\ref{Th_Erg_R} and Theorem~\ref{Th_Erg_R_lowSNR}.

\begin{Theorem}
\label{Th_Erg_R}
For high SNR,
the spatially ergodic rate defined in \eqref{Eq_Er_Exp} with moderate-to-large $N$ is given by
\begin{align}
    \Tilde{R} = &\log\left(\frac{P}{\sigma^2}\beta \right)
    \!+\!\left(1\!-\!\mathrm{e}^{-\pi\lambda C^2}\right)\log\beta
    +\mathcal{H}(N,\mu)
    +\bar{\mathcal{G}}(N,\mu)
    \nonumber\\&
    -\frac{(\alpha_1\!-\!\alpha_2)\mathrm{e}^{-\pi\lambda C^2}+\alpha_2}{\ln 2} \left(\frac{D_2^2\ln D_2 -D_1^2\ln D_1 }{D_2^2-D_1^2} -\frac{1}{2}\!\right)
    \nonumber\\&
    -\frac{\alpha_3}{2\ln 2}
    \left[\mathrm{Ei} \left(-\pi\lambda C^2\right)
    \!-\mathrm{e}^{-\pi\lambda C^2} \ln C^2
    -\ln(\pi\lambda)-E_0\right]
    \label{Eq_Erg_R}
    \!,
\end{align}
where we define
\begin{align}
\mathcal{H}(N,\mu) \triangleq   &
    \log\left[ N \left(\mu^2 N +1-\mu^2\right)\right] \left(1-\mathrm{e}^{-\pi\lambda C^2}\right)
	,
	\label{Eq_H}
\end{align}
\begin{align}
&\bar{\mathcal{G}}(N,\mu) \triangleq
 \frac{1}{\beta\ln 2}\!
\left(\!\frac{ \kappa_1 \sqrt{\pi\beta}\mu\gamma\left(\frac{\alpha_3}{4}\!\!+\!\!1,\pi\lambda C^2\right) }{(\pi\lambda)^{\frac{\alpha_3}{4}} (\mu^2 N \!+\!1\!-\!\mu^2)}
\!\!+\!\!\frac{\kappa_2\gamma\left(\frac{\alpha_3}{2}+1,\pi\lambda C^2\right)}{(\pi\lambda)^{\frac{\alpha_3}{2}}N \left(\mu^2 N\! +\!1\!\!-\!\!\mu^2\right)}  \!\!\right)
\!\!,
\label{Eq_G_bar}
\end{align}
\begin{align}
    \kappa_1 \triangleq \frac{4\left(D_2^\frac{\alpha_2-\alpha_1+4}{2}-D_1^\frac{\alpha_2-\alpha_1+4}{2}\right)}{(\alpha_2-\alpha_1+4)\left(D_2^2-D_1^2\right)}
    \label{ka1}
    ,
\end{align}
and
\begin{align}
    \kappa_2 \triangleq \frac{2\left(D_2^{\alpha_2-\alpha_1+2}-D_1^{\alpha_2-\alpha_1+2}\right)}{(\alpha_2-\alpha_1+2)\left(D_2^2-D_1^2\right)}
    \label{ka2}
    .
\end{align}
\end{Theorem}

\begin{IEEEproof}
See Appendix~\ref{Proof_Th2} with preliminaries in Appendix~\ref{Ap_Useful}.
\end{IEEEproof}

\begin{Theorem}
\label{Th_Erg_R_lowSNR}
For low SNR,
the spatially ergodic rate defined in \eqref{Eq_Er_Exp} with moderate-to-large $N$ is given by
\begin{align}
    \Tilde{R} = &
    \left(1\!-\!\mathrm{e}^{-\pi\lambda C^2}\right)\!\log\!\left(\!\frac{P}{\sigma^2}\beta^2\!\right)
    \!+\!\mathcal{H}(N,\mu)
    \!+\!\bar{\mathcal{G}}(N,\mu)
    \!-\!\left(\!1\!-\!\mathrm{e}^{-\pi\lambda C^2}\!\right)\!\frac{\alpha_2}{\ln 2} \!\left(\!\frac{D_2^2\ln D_2 \!-\!D_1^2\ln D_1 }{D_2^2-D_1^2} \!-\!\frac{1}{2}\!\right)
    \nonumber\\&
    -\!\frac{\alpha_3}{2\ln 2}
    \left[\!\mathrm{Ei} \left(\!-\pi\lambda C^2\!\right)
    \!-\!\mathrm{e}^{-\pi\lambda C^2} \!\ln C^2
    \!-\!\ln(\!\pi\lambda\!)\!-\!E_0\!\right]
    \!\!+\!\!\frac{\mathrm{e}^{-\pi\lambda C^2} }{\ln 2} \frac{P\beta}{\sigma^2}  \frac{2\left(\!D_2^{2-\alpha_1}\!-\!D_1^{2-\alpha_1}\!\right)}{(\!2\!-\!\alpha_1\!)\left(D_2^2\!-\!D_1^2\right)}
    \! +\!\bar{\mathcal{G}}_L(\!N,\mu\!)
    \label{Eq_Erg_R_L}
    ,
\end{align}
where we define
\begin{align}
&\bar{\mathcal{G}}_L(N,\mu) \triangleq
 \frac{1}{\beta\ln 2}\!
	\frac{\kappa_3\gamma\left(\frac{\alpha_3}{2}+1,\pi\lambda C^2\right)}{(P\beta/\sigma^2)(\pi\lambda)^{\frac{\alpha_3}{2}}N \left(\mu^2 N\! +\!1\!-\!\mu^2\right)}
       \!,
\label{Eq_G_bar_L}
\end{align}
and
\begin{align}
    \kappa_3 \triangleq \frac{2\left(D_2^{\alpha_2+2}-D_1^{\alpha_2+2}\right)}{(\alpha_2+2)\left(D_2^2-D_1^2\right)}
    \label{ka3}
    .
\end{align}
\end{Theorem}

\begin{IEEEproof}
See Appendix~\ref{Proof_Th3} with preliminaries in Appendix~\ref{Ap_Useful}.
\end{IEEEproof}

\begin{Observation}
It is noteworthy that $\bar{\mathcal{G}}(N,\mu)$ and $\mathcal{H}(N,\mu)$ are involved in both \eqref{Eq_Erg_R} and \eqref{Eq_Erg_R_L}. We compare these two key terms as follows.
From \eqref{Eq_H} and \eqref{Eq_G_bar}, it is obvious that $\bar{\mathcal{G}}(N,\mu)$ is a decreasing function while $\mathcal{H}(N,\mu)$ is an increasing function w.r.t. $N$.
Consider an upper bound of $\bar{\mathcal{G}}(N,\mu)$ expressed as
\begin{align}
    \bar{\mathcal{G}}(N,\mu)
    =&	\frac{1}{\beta\ln 2}\!\!
	\left(\!\!\frac{ \kappa_1 \sqrt{\pi\beta}\mu\gamma\!\left(\!\frac{\alpha_3}{4}\!\!+\!\!1,\pi\lambda C^2\right) }{(\pi\lambda)^{\frac{\alpha_3}{4}} \left[\mu^2 (N-1) \!+\!1\right]}
	\!\!+\!\!\frac{\kappa_2\gamma\left(\frac{\alpha_3}{2}+1,\pi\lambda C^2\right)}{(\pi\lambda)^{\frac{\alpha_3}{2}}N \left[\mu^2 (N\!\!-\!\!1\!) \!+\!1\right]}  \!\!\right)
	\nonumber\\\leq&
	\frac{1}{\beta\ln 2}\!\!
	\left(\!\frac{ \kappa_1 \sqrt{\pi\beta}\gamma\left(\frac{\alpha_3}{4}+1,\pi\lambda C^2\right) }{2(\pi\lambda)^{\frac{\alpha_3}{4}} \sqrt{N-1}}
	+\frac{\kappa_2\gamma\left(\frac{\alpha_3}{2}+1,\pi\lambda C^2\right)}{(\pi\lambda)^{\frac{\alpha_3}{2}}N}  \!\right)
	\label{Eq_G_7}
	\\\ll&\mathcal{H}(N,\mu)
	\label{Eq_G_8}
	,
\end{align}
where \eqref{Eq_G_7} comes from $\mu\in[0,\frac{\pi}{4}]$, $N>1$, and $x+y\geq2\sqrt{xy}$ for $\forall x,y\geq 0$, and \eqref{Eq_G_8} is obtained by comparing \eqref{Eq_G_7} with \eqref{Eq_H} for moderate-to-large $N$, which is consistent with the deployment of RISs in practice.
From \eqref{Eq_G_8}, $\bar{\mathcal{G}}(N,\mu)$ is negligible compared to the component of $\mathcal{H}(N,\mu)$ and therefore it is safe to drop it in the following analysis on the spatially ergodic rate for both high and low SNRs.
\end{Observation}

\subsection{Optimal Deployment of Distributed RISs}
From \eqref{Eq_Erg_R} and \eqref{Eq_Erg_R_L}, $\Tilde{R}$ can be enhanced by increasing either $\lambda$ or $N$ at the expense of deploying more RIS elements in total.
Due to the constraint of hardware cost in practice, we introduce a fair and tractable constraint of a total number of RIS elements per unit area, that is  $\lambda N \equiv \eta$ for a constant $\eta$.
Under this constraint, increasing $\lambda$ leads to a reduced value of $N$, making the joint effect on $\Tilde{R}$ uncertain, and vice versa.
To the best of our knowledge, quantitative analysis under this reasonable constraint has rarely been considered in literature.

Next, we optimize $\lambda$ for rate maximization as follows
\begin{align}
    &\underset{\lambda}{\mathrm{maximize}}~~\Tilde{R}
    \label{Fun_R}\\
    &\mathrm{subject~to}~~ \lambda N = \eta,
    ~N\in\mathbb{Z^+}.
    \nonumber
\end{align}
In the following, we first consider the case of high SNR and then elaborate on the differences in optimization for low SNR.
Removing the terms independent of both $\lambda$ and $N$ in \eqref{Eq_Erg_R}, the objective function of \eqref{Fun_R} is equivalently rewritten as
\begin{align}
    \mathcal{F}(N,\lambda) =&
    -\frac{\alpha_3}{2\ln 2}
    \!\left[\!\mathrm{Ei} \left(\!-\!\pi\lambda C^2\right)
    \!-\!\mathrm{e}^{-\pi\lambda C^2} \ln C^2
    \!-\!\ln(\pi\lambda)\!\right]
    -\mathrm{e}^{-\pi\lambda C^2}\!\left(\!D\!+\!\log\beta \right)
    \!+\!\mathcal{H}(N,\mu)
    \label{Fun_F}
    ,
\end{align}
where
\begin{align}
    D \triangleq \frac{(\alpha_1-\alpha_2)}{\ln 2} \left(\frac{D_2^2\ln D_2 -D_1^2\ln D_1 }{D_2^2-D_1^2} -\frac{1}{2}\!\right).
\end{align}
Then from \eqref{Fun_R} and \eqref{Fun_F}, the optimal $\lambda$ can be obtained by maximizing $\mathcal{F}(N,\lambda)$ under the constraint of $\lambda N\equiv \eta$, yielding
\begin{align}
    \lambda^* =  \mathrm{arg}~~&\underset{\lambda}{\mathrm{maximize}}~~{\mathcal{F}\left(\frac{\eta}{\lambda},\lambda\right)}
    \label{Lambda_Opt}\\
    &\mathrm{subject~to}~~\frac{\eta}{\lambda}\in\mathbb{Z^+}
    .
    \nonumber
\end{align}
Since $N=\frac{\eta}{\lambda}\in\mathbb{Z^+}$ is a positive integer, we have $\lambda\in(0,\eta]$.
Considering that ${\mathcal{F}\left(\frac{\eta}{\lambda},\lambda\right)}$ in \eqref{Fun_F} is a continuous function w.r.t $\lambda\in(0,\eta]$, the optimal $\lambda^*$ can be acquired by forcing the derivative of ${\mathcal{F}\left(\frac{\eta}{\lambda},\lambda\right)}$ w.r.t. $\lambda$ to zero, or $\lambda^*$ tends to the critical values, that is $\lambda\rightarrow 0 $ or $\lambda=\eta $.
In the following, we characterize the optimal $\lambda^*$ under the conditions of $\rho\in[0,1)$ and $\rho=1$.
Closed-form expressions of the optimal solutions for typical cases are also obtained.

\subsubsection{Bounded Phase Shift Errors with $\rho\in[0,1)$ for High SNR}
In order to evaluate $\lambda^*$ in \eqref{Lambda_Opt} and characterize the impact of phase shift error on the optimization of RIS deployment, we first consider the bounded phase shift error model with $\rho\in[0,1)$.
From  \eqref{Fun_F}, the derivative of ${\mathcal{F}\left(\frac{\eta}{\lambda},\lambda\right)}$ w.r.t. $\lambda$ is
\begin{align}
    \frac{\partial\mathcal{F}}{\partial \lambda} = &
    \frac{1\!-\!\mathrm{e}^{-\pi\lambda C^2}}{\lambda\ln 2} \left(\!\frac{\alpha_3}{2}\!-\!2\!+\!\frac{1-\mu^2}{\mu^2\eta\lambda^{-1}\!+\!1\!-\!\mu^2}\!\right)
    +\!\pi C^2 \mathrm{e}^{-\pi\lambda C^2}\log\!\left[2^D C^{-\alpha_3}\beta\eta\lambda^{-1}\left(\mu^2\eta\lambda^{-1}\!+\!1\!-\!\mu^2\right) \right]
    \nonumber\\ \triangleq &
    \frac{\mathrm{e}^{-\pi\lambda C^2}}{\lambda \ln 2} J(\lambda)
    ,
\end{align}
where $J(\lambda)$ is defined as
\begin{align}
    J(\lambda)\triangleq&
    \pi \lambda C^2\ln\!\left[2^D C^{-\alpha_3}\beta\eta\lambda^{-1}\left(\mu^2\eta\lambda^{-1}\!+\!1\!-\!\mu^2\right) \right]
    +\left(\!\frac{\alpha_3}{2}\!-\!2\!+\!\frac{1-\mu^2}{\mu^2\eta\lambda^{-1}\!+\!1\!-\!\mu^2}\!\right)\left(\mathrm{e}^{\pi\lambda C^2}\!-\!1\right)
    \label{J}
    \!.
\end{align}
Since $\frac{\mathrm{e}^{-\pi\lambda C^2}}{\lambda \ln 2}>0$ for $\forall\lambda\in(0,\infty)$, $\frac{\partial\mathcal{F}}{\partial \lambda}=0$ has the same solution as that of $J(\lambda)=0$.
Usually for a single RIS equipped with a large number of low-cost reflecting elements, it is reasonable to consider $N\gg\frac{1}{\mu^2}-1$ for $\mu\in(0,\frac{\pi}{4}]$. Due to $N=\frac{\eta}{\lambda}$, it yields $\lambda\ll\frac{\mu^2\eta}{1-\mu^2}$ and the derivative of $J(\lambda)$ w.r.t. $\lambda$ is derived as
\begin{align}
    \frac{\partial J}{\partial \lambda}\! = \!
    \pi C^2 \left[\left(\frac{\alpha_3}{2}\!-\!2\right) \mathrm{e}^{\pi\lambda C^2} \!-\!\ln \lambda
    \!+\!\ln\left(2^D C^{-\alpha_3}\beta\mu^2\eta^2\right) \!-\!2 \right]
    \label{J_De}
    \!\!.
\end{align}
From \eqref{J_De}, it is obvious that $\frac{\partial J}{\partial \lambda}$ decreases monotonically w.r.t. $\lambda\in(0,\infty)$ since $\frac{\alpha_3}{2}-2\leq 0$ with $\alpha_3\in[2,4]$.
Also, we have $\frac{\partial J}{\partial \lambda}\rightarrow\infty$ when $\lambda\rightarrow 0 $, while $\frac{\partial J}{\partial \lambda}\rightarrow-\infty$ when $\lambda\rightarrow \infty$.
Hence, the equation $ \frac{\partial J}{\partial \lambda}=0$ has a single unique positive solution, denoted by $\lambda_0$, which indicates that $J(\lambda)$ increases when $\lambda\in(0,\lambda_0)$ while decreases for $\lambda\in(\lambda_0,\infty)$.
Along with $J(0)\rightarrow 0$ and $J(\infty)\rightarrow -\infty$, the equation $ \frac{\partial F}{\partial \lambda}=J(\lambda)=0$ has a single unique positive solution, denoted by $\lambda_1$, yielding that
$\frac{\partial\mathcal{F}}{\partial \lambda}=\frac{\mathrm{e}^{-\pi\lambda C^2}}{\lambda \ln 2} J(\lambda)>0$ for $\lambda\in(0,\lambda_1)$ and $\frac{\partial\mathcal{F}}{\partial \lambda}=\frac{\mathrm{e}^{-\pi\lambda C^2}}{\lambda \ln 2} J(\lambda)<0$ for $\lambda\in(\lambda_1,\infty)$.
With the fact of $\lambda\in(0,\eta]$, we conclude that
\begin{align}
    \lambda^*=\min\{\lambda_1,\eta\},
\end{align}
is exactly the solution to problem \eqref{Lambda_Opt}.
Although it is hard to acquire a closed-form expression of $\lambda_1$, we can resort to efficient numerical methods, e.g., the bisection search method.
Correspondingly, the optimal $N$ is obtained as
\begin{align}
    N^*=\left\lceil\frac{\eta}{\lambda^*}\right\rceil.
\end{align}
It is noteworthy that for a typical case of $\alpha_3=4$, closed-form expressions of $\lambda^*$ and $N^*$ are fortunately available in the following Lemma.

\begin{Lemma}
\label{lem_opt}
Under the condition of imperfect phase shifts with $\rho\in[0,1)$ and $\alpha_3=4$, the optimal choices of $\lambda$ and $N$ under the constraint of a total number of RIS elements per unit area, i.e., $\lambda N \equiv \eta$, to maximize the spatially ergodic rate in \eqref{Eq_Erg_R} are, respectively,
\begin{align}
    \lambda^* = \min\left\{ \mu\eta C^{-2}\sqrt{2^D \beta},~\eta\right\},
    \label{lambda_opt}
\end{align}
and
\begin{align}
    N^* = \left\lceil \mu^{-1} C^{2} \sqrt{2^{-D}\beta^{-1} }\right\rceil.
    \label{N_opt}
    \end{align}
\end{Lemma}
\begin{IEEEproof}
Substituting $\alpha_3=4$ into \eqref{J} and using $\lambda\ll\frac{\mu^2\eta}{1-\mu^2}$, $J(\lambda)$ is simplified to
\begin{align}
    J(\lambda)=
    \pi\lambda C^2\ln\left(2^D C^{-4}\beta\mu^2\eta^2\lambda^{-2}\right)
    .
\end{align}
Then, the solution to $J(\lambda)=0$ is $\lambda_1=\mu\eta C^{-2}\sqrt{2^D}\beta$ and $\lambda^*$ in \eqref{lambda_opt} is obtained.
Correspondingly, the optimal $N^*$ is obtained by substituting \eqref{lambda_opt} into $N^*=\left\lceil\frac{\eta}{\lambda^*}\right\rceil$.
\end{IEEEproof}

\begin{Remark}
\label{Re_OptN_1}
From \eqref{N_opt}, we observe that \emph{the optimal $N$ increases quadratically w.r.t. $C$ while decreases w.r.t. $\mu$.}
Under the constraint of a total number of RIS elements per unit area, \emph{it is suggested to form larger RISs with sparser density to support larger coverage and combat severer phase shift errors.}
Besides, a smaller $\beta$, corresponding to severer large-scale fading of the cascaded BS-RIS-UE channel, also requires a larger $N^*$.
This observation is also valid for general cases with $\alpha_3\in[2,4)$ as verified by numerical results in Section~\ref{Sec_Sim}.
\end{Remark}

\subsubsection{Random Phase Shifts with $\rho=1$ for High SNR}
In order to evaluate the optimal $\lambda^*$ for random phase shifts, i.e., $\rho=1$, we substitute $\mu=0$ into \eqref{J}. It yields
\begin{align}
    J(\lambda)\triangleq
    \left(\!\frac{\alpha_3}{2}\!-1\right)\left(\mathrm{e}^{\pi\lambda C^2}\!-\!1\right)
    \!+\!\pi \lambda C^2\ln\!\left[2^D C^{-\alpha_3}\beta\eta\lambda^{-1} \right]
    \label{J_mu0}
    ,
\end{align}
and
\begin{align}
    \frac{\partial J}{\partial \lambda} =
    \pi C^2 \left[\left(\frac{\alpha_3}{2}-1\right) \mathrm{e}^{\pi\lambda C^2} -\ln \lambda
    +\ln\left(2^D C^{-\alpha_3}\beta\eta\right) -1 \right]
    \!\!.
    \label{J_De_mu0}
\end{align}
Different from \eqref{J_De}, $\frac{\partial J}{\partial \lambda}$ in \eqref{J_De_mu0} is in general not a monotonous function except for the special case of $\alpha_3=2$.
In this typical case, the closed-form expressions of $\lambda^*$ and $N^*$ are presented in Lemma~\ref{lem_opt2}.

\begin{Lemma}
\label{lem_opt2}
Under the condition of random phase shifts with $\rho=1$ and $\alpha_3=2$, the optimal choices of $\lambda$ and $N$ under the constraint of a total number of RIS elements per unit area, i.e., $\lambda N \equiv \eta$, to maximize the spatially ergodic rate in \eqref{Eq_Erg_R} are, respectively, given as
\begin{align}
    \lambda^*=\min\left\{2^D C^{-2}\beta\eta,\eta\right\},
    \label{lambda_opt_2}
\end{align}
and
\begin{align}
    N^*=\left\lceil 2^{-D} C^{2}\beta^{-1} \right\rceil.
    \label{N_opt_2}
\end{align}
\end{Lemma}
\begin{IEEEproof}
Substituting $\alpha_3=2$ into \eqref{J_mu0}, it yields
\begin{align}
    J (\lambda) = \pi \lambda C^2 \log\!\left(2^D C^{-2}\beta\eta\lambda^{-1} \right),
    \label{J_mu0_2}
\end{align}
which decreases w.r.t. $\lambda\in(0,\infty)$.
By forcing $J (\lambda) =0$, we denote the solution by $\lambda_3 = 2^D C^{-2}\beta\eta$, yielding $\frac{\partial\mathcal{F}}{\partial \lambda}=\frac{\mathrm{e}^{-\pi\lambda C^2}}{\lambda \ln 2} J(\lambda)>0$ for $\lambda\in(0,\lambda_3)$ and $\frac{\partial\mathcal{F}}{\partial \lambda}=\frac{\mathrm{e}^{-\pi\lambda C^2}}{\lambda \ln 2} J(\lambda)<0$ for $\lambda\in(\lambda_3,\infty)$.
Due to the fact $\lambda\in(0,\eta]$ and $N^*=\left\lceil\frac{\eta}{\lambda^*}\right\rceil$, we obtain the optimal $\lambda^*$ and $N^*$ that to maximize $\Tilde{R}$ in \eqref{lambda_opt_2} and \eqref{N_opt_2}, respectively.
\end{IEEEproof}

For a general case of $\alpha_3\in(2,4]$, there are unfortunately no closed-form solutions to $J(\lambda)=0$ while $\Tilde{R}$ still increases monotonically w.r.t. $\lambda\in(0,\eta]$. The optimal solutions under this condition are presented in Lemma~\ref{lem_opt3}.
\begin{Lemma}
\label{lem_opt3}
Under the conditions of random phase shifts with $\rho=1$ and $\alpha_3\in(2,4]$, the optimal $\lambda$ and $N$ under the constraint of a large limited number of RIS elements per unit area, i.e., $\lambda N \equiv \eta\geq\frac{2C^{\alpha_3-2}}{(\alpha_3-2)\pi e \beta 2^D}$,
to maximize the spatially ergodic rate in \eqref{Eq_Erg_R} are, respectively,
\begin{align}
    \lambda^* = \eta,
    \label{lambda_opt_3}
\end{align}
and
\begin{align}
    N^* = 1.
    \label{N_opt_3}
\end{align}
\end{Lemma}
\begin{IEEEproof}
From \eqref{J_De_mu0}, the second derivative of $J(\lambda)$ w.r.t. $\lambda$ is expressed as
\begin{align}
    \frac{\partial^2 J}{\partial \lambda^2} =
    \pi C^2 \left[\pi C^2 \left(\frac{\alpha_3}{2}-1\right) \mathrm{e}^{\pi\lambda C^2} -\frac{1}{\lambda} \right]
    .
\end{align}
It is observed that $\frac{\partial^2 J}{\partial \lambda^2}$ increases w.r.t. $\lambda\in(0,\infty)$ and the equation $\frac{\partial^2 J}{\partial \lambda^2}=0$ has only a single positive solution, say $\lambda_2$.
By substituting $\lambda_2=\left[\pi C^2 \left(\frac{\alpha_3}{2}-1\right) \mathrm{e}^{\pi\lambda_2 C^2}\right]^{-1}$ into \eqref{J_De_mu0}, the minimum of $\frac{\partial J}{\partial \lambda}$ satisfies
\begin{align}
    \frac{\partial J}{\partial \lambda} \bigg|_{\lambda_2}
    =    &\pi C^2 \!\!\left[\!\frac{1}{\pi\lambda_2 C^2}\!\!+\! \pi\lambda_2 C^2
    \!\!+ \!\ln\!\left[\!\pi C^2\! \left(\!\frac{\alpha_3}{2}\!\!-\!\!1\!\right) \!\right]
    \!\!+\!\!\ln\left(2^D C^{-\alpha_3}\beta\eta\right)\!\! -\!\!1\!\! \right]
    \nonumber\\\geq&
    \pi C^2
    \ln\left[\pi e \beta\eta 2^D C^{2-\alpha_3}\left(\frac{\alpha_3}{2}-1\right) \right]
    \\\geq &0
    ,
\end{align}
where we use the inequality $x+y\geq 2\sqrt{xy}$ for $\forall x,y\geq 0$ and the assumption $\eta\geq\frac{2C^{\alpha_3-2}}{(\alpha_3-2)\pi e \beta 2^D}$.
Then, according to $\frac{\partial J}{\partial \lambda} \big|_{\forall\lambda\in(0,\infty)} \geq \frac{\partial J}{\partial \lambda} \big|_{\lambda_2} \geq 0$ and $J(0)=0$, we have $\frac{\partial\mathcal{F}}{\partial \lambda}=\frac{\mathrm{e}^{-\pi\lambda C^2}}{\lambda \ln 2} J(\lambda)\geq0$ for $\lambda\in(0,\eta]$.
Therefore, $\Tilde{R}$ increase w.r.t. $\lambda\in(0,\eta]$ and it reaches the maximum at $\lambda^*=\eta$ and $N^*=1$.
\end{IEEEproof}

Under the constraint of $\lambda N \equiv \eta$, a higher $\lambda$ statistically shortens the distance between the UE and its nearest RIS, meanwhile the resulting smaller $N$ reduces the beamforming gain of the RIS.
Obviously, there is a deployment tradeoff between the choices of parameters $\lambda$ and $N$ in a multi-RIS-aided network.
However, $\Tilde{R}$ increases monotonically w.r.t. $\lambda$ under the conditions of $\rho=1$ and $\alpha_3\in(2,4]$ as indicated in Lemma~\ref{lem_opt3}.
This is because as $\lambda$ increases, the improved channel gain corresponding to the statistically shorter RIS-UE distance overwhelms the loss of beamforming gain due to smaller RISs. 
In this typical case, a higher RIS deployment density is proven always beneficial to the network performance.

Next, we consider the case of low SNR using Theorem~\ref{Th_Erg_R_lowSNR}.
Although the overall derivations for the optimization problem in \eqref{Fun_R} for low SNR are similar to that in the case of high SNR, a small value of $\frac{P}{\sigma^2}$ introduces additional terms that are no longer negligible, which consequently makes it intractable to get a closed-form expression for the optimal $\lambda^*$, as obtained in Lemmas~\ref{lem_opt}-\ref{lem_opt2} under high SNR.
To be specific, we remove the terms independent of both $\lambda$ and $N$ in \eqref{Eq_Erg_R_L}, yielding the objective function for \eqref{Lambda_Opt} as
\begin{align}
    \mathcal{F}(N,\lambda) =&
     -\mathrm{e}^{-\pi\lambda C^2}D
    \!+\!\mathcal{H}(N,\mu)+\bar{\mathcal{G}}_L(N,\mu)
    -\frac{\alpha_3}{2\ln 2}
    \!\left[\!\mathrm{Ei} \left(\!-\!\pi\lambda C^2\right)
    \!-\!\mathrm{e}^{-\pi\lambda C^2} \ln C^2
    \!-\!\ln(\pi\lambda)\!\right]
    \label{Fun_F_L}
    ,
\end{align}
where $D$ is independent of $\lambda$ and $N$, defined as
\begin{align}
    D\triangleq &\log\left(\frac{P}{\sigma^2}\beta^2\right)
    -\frac{\alpha_2}{\ln 2} \left(\frac{D_2^2\ln D_2 -D_1^2\ln D_1 }{D_2^2-D_1^2} -\frac{1}{2}\!\right)
    -  \frac{P\beta}{\sigma^2\ln 2}  \frac{2\left(D_2^{2-\alpha_1}-D_1^{2-\alpha_1}\right)}{(2-\alpha_1)\left(D_2^2-D_1^2\right)}.
    \label{Eq_D_L}
\end{align}
Then, similar to the above demonstrations for high SNR, we characterize the derivative of ${\mathcal{F}\left(\frac{\eta}{\lambda},\lambda\right)}$ in \eqref{Fun_F_L} w.r.t. $\lambda$ as $\frac{\partial\mathcal{F}}{\partial \lambda} = \frac{\mathrm{e}^{-\pi\lambda C^2}}{\lambda \ln 2} J(\lambda)$, and find the optimal $\lambda^*$ by forcing $J(\lambda)=0$.
In the following, we derive $J(\lambda)$ under the conditions of $\rho\in[0,1)$ and $\rho=1$ separately.

\subsubsection{Bounded Phase Shift Errors with $\rho\in[0,1)$ for Low SNR}
Under the assumption of moderate-to-large $N$, we have $\frac{\partial\mathcal{F}}{\partial \lambda} = \frac{\mathrm{e}^{-\pi\lambda C^2}}{\lambda \ln 2} J(\lambda)$
where $J(\lambda)$ is
\begin{align}
    J(\lambda)=&
    \pi \lambda C^2\ln\!\left[2^{D} C^{-\alpha_3} \mu^2\eta^2\lambda^{-2} \right]
    +\left(\!\frac{\alpha_3}{2}\!-\!2\!\right)\left(\mathrm{e}^{\pi\lambda C^2}\!-\!1\right)
    \nonumber\\&
    +\!\left[  \left(\pi \lambda C^2\right)^{\frac{\alpha_3}{2}+1}\!\! +\! \mathrm{e}^{\pi\lambda C^2} \!\!\left(\!2\!-\!\frac{\alpha_3}{2}\right)\gamma\left(\!\frac{\alpha_3}{2}\!+\!1,\pi\lambda C^2\!\right) \right]
    \frac{\kappa_3 \lambda^{2-\frac{\alpha_3}{2}}}{\frac{P}{\sigma^2}\beta^2  \pi^{\frac{\alpha_3}{2}} \mu^2 \eta^2 }
    \label{J_L}
    \!.
\end{align}
Different from the case of high SNR, the third term in \eqref{J_L} involves the small value of $\frac{P}{\sigma^2}$ and the closed-form solution to the equation $J(\lambda)=0$ is intractable, even for the special case of $\alpha_3=2$.
Alternatively, we resort to an efficient bisection search method.

\subsubsection{Random Phase Shifts with $\rho=1$ for Low SNR}
By substituting $\mu=0$ into \eqref{Fun_F_L}, we have $\frac{\partial\mathcal{F}}{\partial \lambda} = \frac{\mathrm{e}^{-\pi\lambda C^2}}{\lambda \ln 2} J(\lambda)$
where $J(\lambda)$ is
\begin{align}
    J(\lambda)=&
    \pi \lambda C^2\ln\!\left[2^{D} C^{-\alpha_3} \eta\lambda^{-1} \right]
    +\left(\!\frac{\alpha_3}{2}\!-\!1\!\right)\left(\mathrm{e}^{\pi\lambda C^2}\!-\!1\right)
    \nonumber\\&
    +\!\left[\left(\pi \lambda C^2\right)^{\frac{\alpha_3}{2}+1}\!\! +\! \mathrm{e}^{\pi\lambda C^2} \!\left(\!1\!-\!\frac{\alpha_3}{2}\!\right)
    \gamma\left(\!\frac{\alpha_3}{2}\!+\!1,\pi\lambda C^2\!\right)\! \right]
    \frac{\kappa_3 \lambda^{1-\frac{\alpha_3}{2}}}{\frac{P}{\sigma^2}\beta^2  \pi^{\frac{\alpha_3}{2}} \eta }
    \label{J_L_Random}
    \!.
\end{align}
For a general case with $\alpha_3\in(2, 4]$ yielding $\left(\!\frac{\alpha_3}{2}\!-\!1\!\right)>0$,
the optimal parameters are obtained the same as in \eqref{lambda_opt_3} and \eqref{N_opt_3},
since ${\mathcal{F}\left(\frac{\eta}{\lambda},\lambda\right)}$ increases monotonically w.r.t. $\lambda\in(0,\eta]$ with
 $\frac{\partial\mathcal{F}}{\partial \lambda}=\frac{\mathrm{e}^{-\pi\lambda C^2}}{\lambda \ln 2} J(\lambda)\geq0$.
The demonstration follows a similar mathematical derivation to the proof of Lemma~\ref{lem_opt3} and is omitted here due to space limit.

In particular for the special case of $\alpha_3=2$, $J(\lambda)$ in  \eqref{J_L_Random} becomes
\begin{align}
    J(\lambda)=&
    \pi \lambda C^2
    \left[\ln\!\left(2^{D} C^{-2} \eta \right) -\ln\lambda
    +\frac{\kappa_3 \pi \lambda C^2}{\frac{P}{\sigma^2}\beta^2  \pi \eta } \right]
    \label{J_L_Random_2}
    .
\end{align}
Compared to \eqref{J_mu0_2} for high SNR, \eqref{J_L_Random_2} comprises an additional term involving $\frac{P}{\sigma^2}$.
By forcing $J(\lambda)=0$, this equation can also be solved by bisection search, the results of which are validated by numerical results in Section~V.

\subsection{Impact of Phase Shift Errors}
The imperfect phase shifts inevitably leads to the degradation of spatially ergodic rate.
Using the derived rate expression given in Theorem~\ref{Th_Erg_R}, we characterize the ergodic rate loss due to the phase shift errors at high SNR.
While for low SNR, thermal noise dominates the rate performance and the effect of phase shift errors is not evident.
From \eqref{Eq_Erg_R}, the impact of $\mu$ on $\Tilde{R}$ is dominated by the term of $\mathcal{H}(N,\mu)$, which is defined in \eqref{Eq_H}.
Compared to the ideal phase shifts with $\mu=\frac{\pi}{4}$, the rate loss caused by the imperfect phase shifts with $\mu\in\left[0,\frac{\pi}{4}\right)$ is characterized by
\begin{align}
    \Delta \Tilde{R} (N,\mu) \triangleq &
    \mathcal{H}\left(N,\frac{\pi}{4}\right)- \mathcal{H}\left(N,\mu\right)
    \nonumber\\ = &
    \left(1-\mathrm{e}^{-\pi\lambda C^2}\right)
    \log\frac{\frac{\pi^2}{16} N +1-\frac{\pi^2}{16}} {\mu^2 N +1-\mu^2}
    .\label{Eq_Rloss}
\end{align}
Random phase shifts without any knowledge of CSI statistically corresponds to the maximum rate loss, as characterized in the following lemma.

\begin{Lemma}
Assuming that the RIS uniformly selects phase shifts within $\mathcal{U}[-\pi, \pi]$ without the knowledge of CSI, i.e., $\mu=\frac{\sin\left(\rho\pi\right)}{4\rho}\big|_{\rho=1}=0$, the rate loss compared to the implementation of ideal phase shifts follows
\begin{align}
    \Delta \Tilde{R} (N,0)
    = \left(1-\mathrm{e}^{-\pi\lambda C^2}\right)\log \left(\frac{\pi^2}{16} N +1-\frac{\pi^2}{16}\right)
    \sim\log N
    \label{Eq_ErgR_Loss_a1}
    .
\end{align}
\end{Lemma}
\begin{IEEEproof}
The rate loss in \eqref{Eq_ErgR_Loss_a1} is obtained by substituting $\mu=0$ into \eqref{Eq_Rloss}.
\end{IEEEproof}

From \eqref{Eq_ErgR_Loss_a1}, it is evident that the rate loss caused by the random phase shifts scales up as $\log N$.
In addition to the typical case of $\mu=0$, we discuss a general case with $\mu\in\left(0,\frac{\pi}{4}\right]$ in Lemma~\ref{lemma_ErgRLoss}.

\begin{Lemma}
\label{lemma_ErgRLoss}
For large $N\gg\frac{1}{\mu^2}-1$ with $\mu\in\left(0,\frac{\pi}{4}\right]$, the rate loss caused by the phase shift error at the RIS asymptotically converges to a constant as
\begin{align}
    \Delta \Tilde{R} (N,\mu)
    \rightarrow\left(1-\mathrm{e}^{-\pi\lambda C^2}\right)\log \frac{\pi^2}{16\mu^2}
    \label{Eq_ErgR_Loss_1}
    .
\end{align}
While for moderate $N\ll\frac{1}{\mu^2}-1$ with $\mu\in\left(0,\frac{\pi}{4}\right]$, the rate loss increases monotonically w.r.t $N$ as
\begin{align}
    \Delta \Tilde{R} (N,\mu)
    \rightarrow&
    \left(1-\mathrm{e}^{-\pi\lambda C^2}\right)
    \log\left(\frac{\pi^2}{16} N +1-\frac{\pi^2}{16}\right)
    \label{Eq_ErgR_Loss_2}
    ,
\end{align}
which is the same as the case of $\mu=0$ in \eqref{Eq_ErgR_Loss_a1}.
\end{Lemma}

\begin{IEEEproof}
Under the conditions of $N\gg\frac{1}{\mu^2}-1$ and $\mu\in\left(0,\frac{\pi}{4}\right]$, it yields $\frac{\pi^2}{16}N^2\geq\mu^2 N\gg 1-\mu^2\geq 1-\frac{\pi^2}{16}$.
Thus, the rate loss in \eqref{Eq_ErgR_Loss_1} is obtained from \eqref{Eq_Rloss}.
While under the condition of $N\ll\frac{1}{\mu^2}-1$, it yields $\mu^2\ll \frac{1}{N+1}<1$. Substituting this inequality into \eqref{Eq_Rloss}, the rate loss in \eqref{Eq_ErgR_Loss_2} is obtained.
\end{IEEEproof}

\begin{Remark}
For an RIS with a growing size, i.e., $N\rightarrow\infty$, the rate loss caused by random phase shifts with $\rho=1$ tends to infinity, as indicated in \eqref{Eq_ErgR_Loss_a1}.
However, the rate loss in \eqref{Eq_ErgR_Loss_1} saturates as $N\rightarrow\infty$, given a bounded range of phase error with $\rho<1$.
Since $\Delta \Tilde{R}$ in \eqref{Eq_Rloss} increases with $N$, the rate loss is upper bounded by \eqref{Eq_ErgR_Loss_1} for $\rho<1$.
For example, considering a typical case with $\rho=\frac{1}{2}$ yielding  $\mu = \frac{1}{2}$,
the rate loss in \eqref{Eq_ErgR_Loss_1} for large $N$ is upper bounded by
\begin{align}
    \Delta \Tilde{R}\left(N,\frac{1}{2}\right)\! \rightarrow\!
    \left(1\!-\!\mathrm{e}^{-\pi\lambda C^2}\right)\log\frac{\pi^2}{4}<1.30~\mathrm{bps/Hz}.
\end{align}
It implies that \emph{the rate loss accounting for imperfect phase shifts is limited even for an extremely large RIS except for the case of random phase shifts.}
\end{Remark}

\section{Simulation Results}
\label{Sec_Sim}
In this section, we verify the derived ergodic rates in \eqref{Eq_R1_up}, \eqref{Eq_R2_up}, \eqref{Eq_Erg_R}, and \eqref{Eq_Erg_R_L}, along with the corresponding theoretical observations via numerical simulations.
In the following, the thermal noise power is set as $\sigma^2=-80$~dBm,
the reference large-scale fading coefficient is $\beta=-30$~dB,
and the carrier frequency is 3.5~GHz chosen from the commonly used sub-6~GHz band for the 5G network.
The path loss exponents are set as $\alpha_1=3$ for the BS-UE channel, $\alpha_2=2$ for the BS-RIS channel, and $\alpha_3=2.5$ for the RIS-UE channel, unless otherwise specified.
The differences in these path loss exponents come from the distinct propagation environments encountered in practice.
For example, an indoor UE is usually located in a rich scattering environment where the corresponding channels comprise numerous non-line-of-sight (NLoS) paths.
Conversely, the BS-RIS channel is usually dominated by a LoS path \cite{QWu_JointBeamf}.

\subsection{Ergodic Rate Given Fixed Locations}

\begin{figure}[tb]
\centering\includegraphics[width=0.6\textwidth]{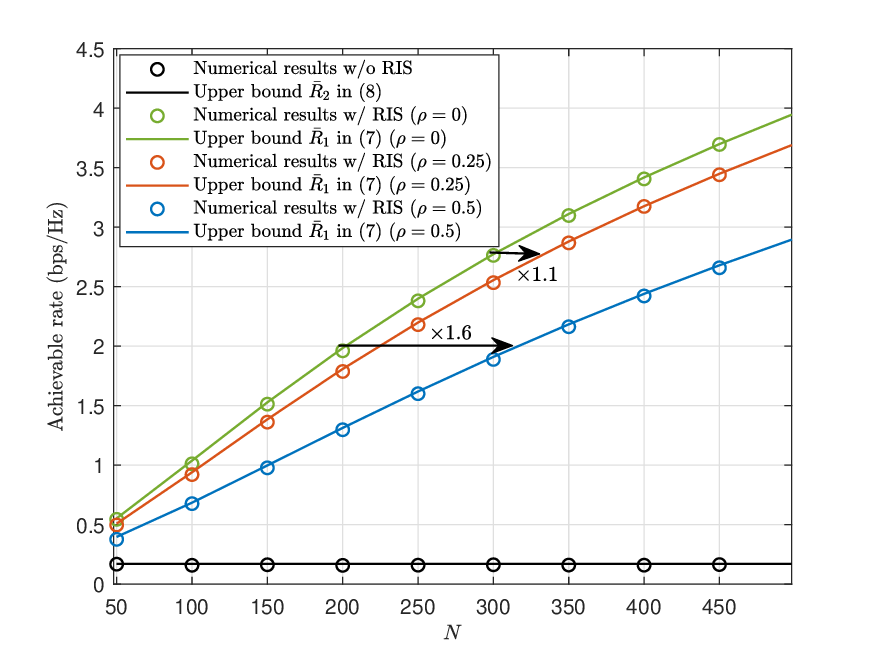}
\caption{Ergodic rate versus $N$ with $P=10$~dBm.}
\label{Fig_R_N}
\end{figure}
\begin{figure}[tb]
\centering\includegraphics[width=0.6\textwidth]{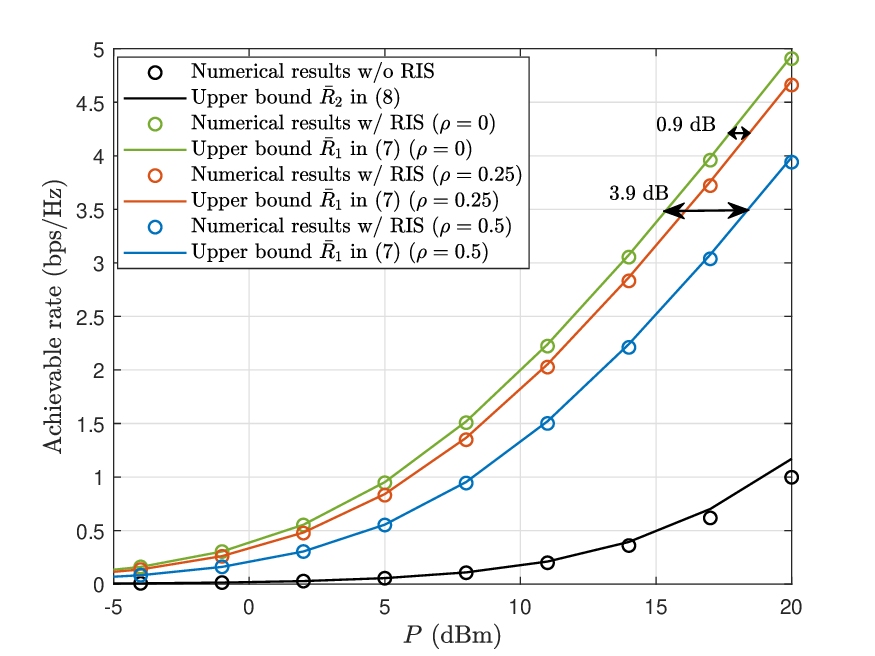}
\caption{Ergodic rate versus $P$ with $N=200$.}
\label{Fig_R_SNR}
\end{figure}
\begin{figure}[tb]
\centering\includegraphics[width=0.6\textwidth]{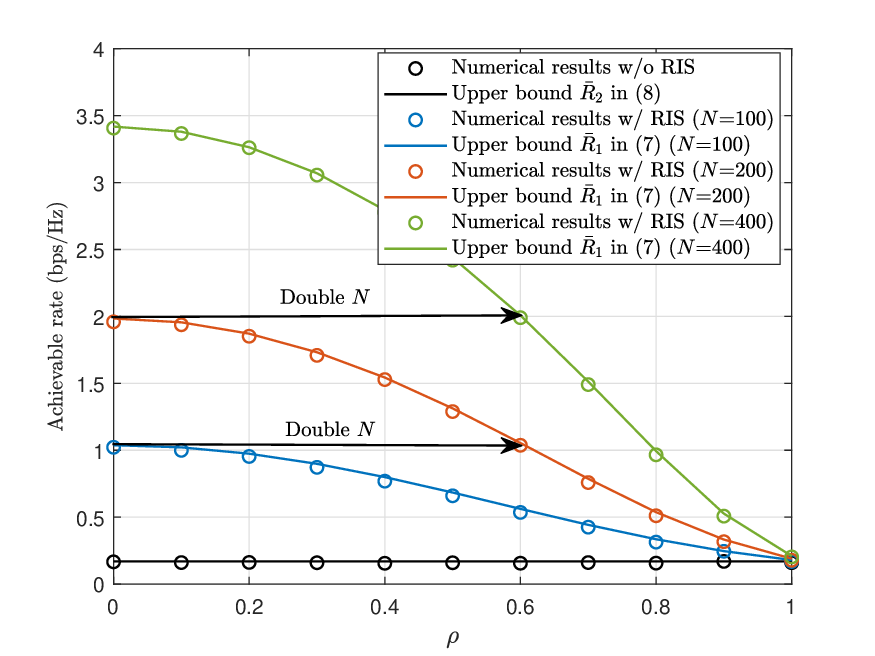}
\caption{Ergodic rate versus $\rho$ with $P=10$~dBm.}
\label{Fig_R_a}
\end{figure}

Given fixed RIS and UE locations with $d=l=200$~m and $r=10$~m, the achievable rates with and without the assistance of an RIS are compared in Figs.~\ref{Fig_R_N}--\ref{Fig_R_a}.
Both the derived rate bounds in \eqref{Eq_R1_up} and \eqref{Eq_R2_up} are verified tight under the tested system parameters.
The comparison results demonstrate the prominent contribution of RIS to the rate enhancement, even if the RIS is suffering from phase shift errors.

In Fig.~\ref{Fig_R_N}, $\bar{R}_1$ increases w.r.t. $N$ for $\rho=0$, $0.25$, and $0.5$, where the increasing slope decreases w.r.t. $\rho$.
This implies that the benefit of large RIS is counteracted by the imperfect phase shifts.
To guarantee that $\bar{R}_1=2$~bps/Hz, the RIS under $\rho=0.5$, i.e., equipped with 1-bit coarse phase shifters, requires about 1.6 times more reflecting elements than the RIS with ideal phase shifts.
Similarly, an extra number of 1.1 times RIS elements can compensate for the rate attenuation due to 2-bit phase shifters with $\rho=0.25$, which is consistent with Remark~\ref{Re_R1}.

From Fig.~\ref{Fig_R_SNR}, the gaps among the ergodic rates with $\rho=0$, $0.25$, and $0.5$ are not evident at low SNRs because thermal noise dominates under this condition.
While at high SNRs, the effect of phase shift errors tends to dominate the performance.
Compared to the ideal benchmark with infinite phase shift resolution, the rate loss resulting from 1-bit and 2-bit discrete phase shifters can be compensated by enlarging the transmit power by $3.9$ and $0.9$~dB, respectively, which is as expected in Remark~\ref{Re_R1}.

Fig.~\ref{Fig_R_a} displays the ergodic rate versus $\rho\in[0,1]$.
It is observed that $\bar{R}_1$ decreases slightly with small $\rho$ while drops sharply as $\rho$ grows.
Obviously, the rate loss caused by subtle phase shift errors is marginal.
It implies that moderate-to-high resolution phase shifters are acceptable in practice to reap most of the benefits of RIS.
However, under the condition of severe phase shift errors, the beamforming gain of RIS diminishes fastly.
Besides, we observe that doubling $N$ is able to overcome the rate attenuation due to the phase shift errors with $\rho=0.6$.

\subsection{Spatially Ergodic Rate Averaged over Random Locations}
\label{Sec_Sim_2}
In the following, we show the numerical results of the spatially ergocic rate averaged over the random locations of the distributed RISs and UEs.
Assuming that the UEs are uniformly located at the cell edge, the distance between UEs and BS is denoted by $d\in[D_1, D_2]$, where $D_1=180$~m and $D_2=220$~m.


\begin{figure}[tb]
\centering\includegraphics[width=0.6\textwidth]{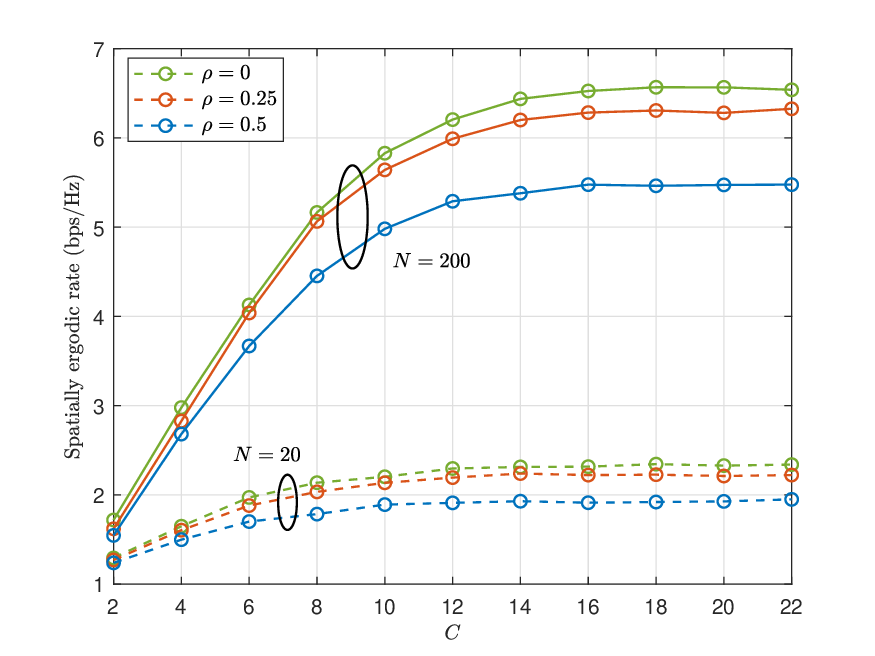}
\caption{Spatially ergodic rate versus $C$ with $P=20$~dBm and $\lambda=0.005$.}
\label{Fig_ErgRate_C}
\end{figure}
\begin{figure}[tb]
\centering\includegraphics[width=0.6\textwidth]{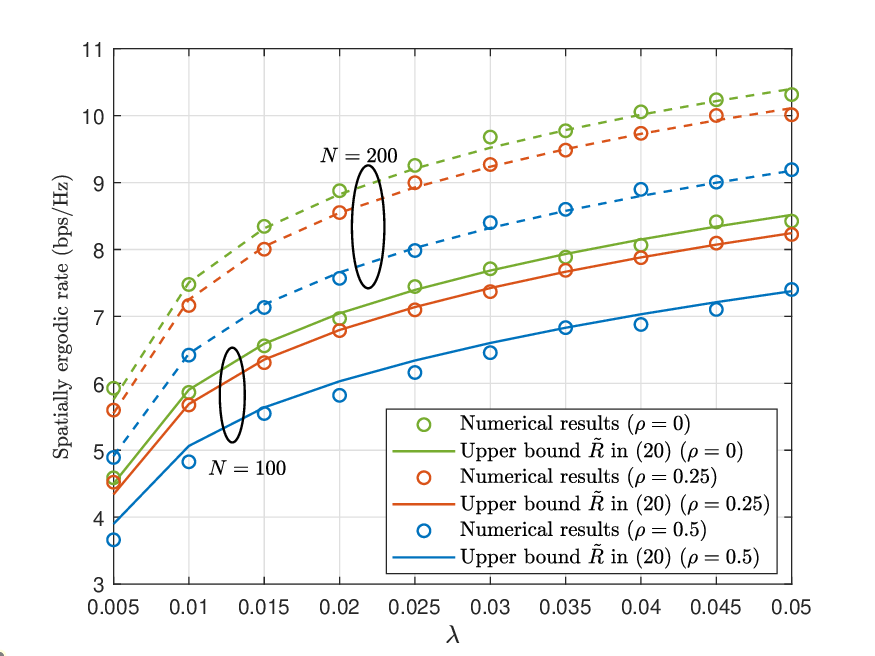}
\caption{Spatially ergodic rate versus $\lambda$ with $P=20$~dBm and $C=10$~m.}
\label{Fig_ErgRate_Lambda}
\end{figure}

Fig.~\ref{Fig_ErgRate_C} displays the spatially ergodic rate versus the RIS serving radius.
It is observed that $\Tilde{R}$ first increases w.r.t. $C$ because a larger serving radius leads to a higher probability of RIS association.
According to the PDF of $r$ in \eqref{PDF_r} with density $\lambda=0.005$, the probabilities of $r<12$ and $r<16$ are 89.6\% and 98.2\%, respectively.
For $N=200$, $\Tilde{R}$ becomes saturated when $C>16$ since the probability of $r\leq C$ can hardly be further enhanced by larger $C$.
While for small $N$ such as $20$, the RIS beamforming gain is extremely marginal due to the severe large-scale fading of the RIS-UE channel with a distance $r>12$, and hence leading to the saturated $\Tilde{R}$ when $C>12$.
It is noteworthy that $C$ is supposed to be shorter than the saturation point in practice, if the additional overhead, including RIS connection, channel estimation, etc, are considered.

Fig.~\ref{Fig_ErgRate_Lambda} shows the spatially ergodic rate versus the deployment density of the distributed RISs.
Given $N$, the ergodic rate increases monotonically w.r.t. $\lambda$ because a higher RIS deployment density statistically squeezes the distance between the RIS and UE.
Given $\lambda$, larger $N$ carries out stronger beamforming and thus improves the ergodic rate.
Nevertheless, the performance gain brought by either higher $\lambda$ or larger $N$ is at the expense of higher hardware cost because totally more RIS elements are required.
Compared to the case with $\lambda=0.01$ and $N=100$, doubling $\lambda$ improves the rate by 1.1~bps/Hz while doubling $N$ improves the rate by 1.6~bps/Hz. Both at the cost of exploiting double RIS elements, the latter choice is more cost-efficient under this condition.

\begin{figure}[tb]
\centering\includegraphics[width=0.6\textwidth]{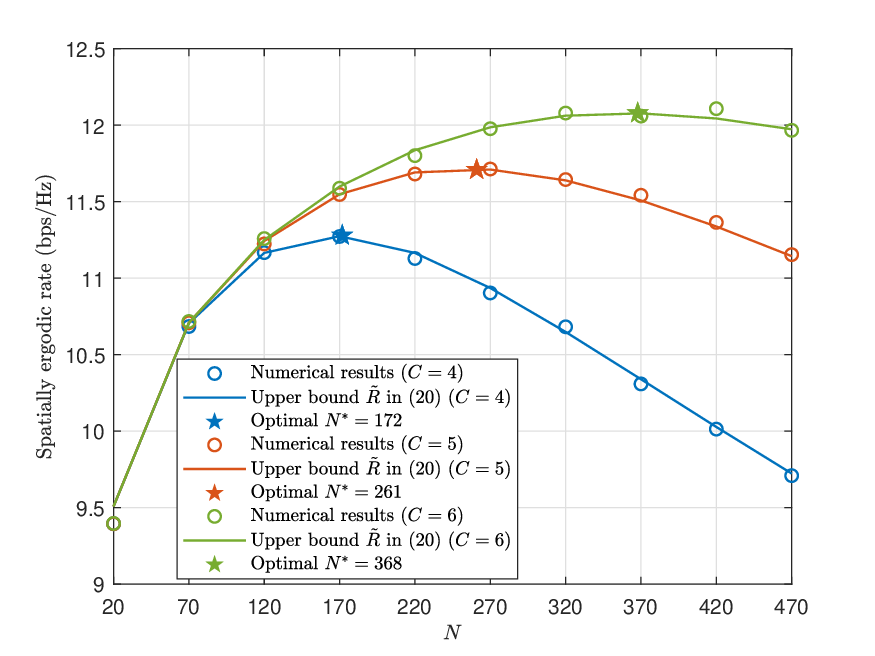}
\caption{Spatially ergodic rate under the constraint of $\eta=10$ with $\rho=0.25$ and $P=25$~dBm.}
\label{Fig_ErgRate_Nopt}
\end{figure}

\begin{figure}[tb]
\centering\includegraphics[width=0.6\textwidth]{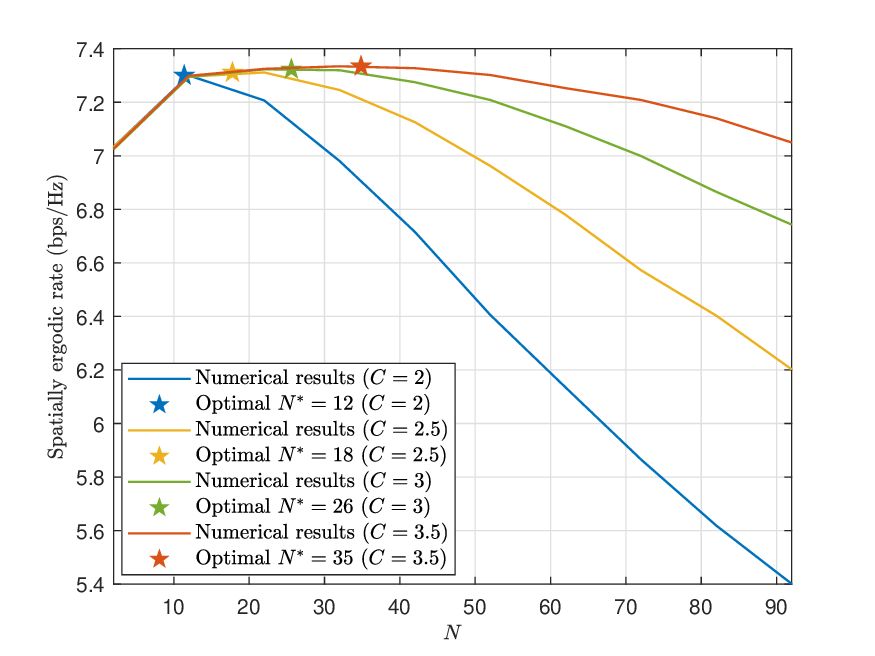}
\caption{Spatially ergodic rate under the constraint of $\eta=5$ with $\rho=0$, $\alpha_3=4$, and $P=20$~dBm.}
\label{Fig_ErgRate_Nopt_typ}
\end{figure}

\begin{figure}[tb]
\centering\includegraphics[width=0.6\textwidth]{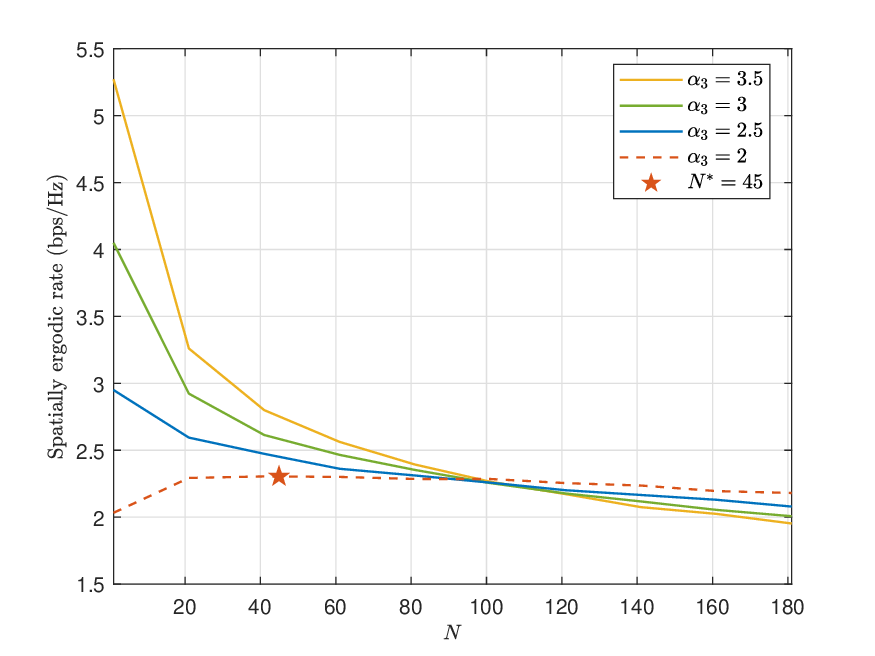}
\caption{Spatially ergodic rate with random phase shifts and constrained by $\lambda N \equiv 10$ ($\rho=1$, $C=3$~m, and $P=15$~dBm).}
\label{Fig_ErgRate_Nopt_rho1}
\end{figure}

Fig.~\ref{Fig_ErgRate_Nopt} and Fig.~\ref{Fig_ErgRate_Nopt_typ} display the spatially ergodic rate versus $N$ under the constraint of $\lambda N \equiv 10$ and $\lambda N \equiv 5$ for fair comparison, respectively.
We observe that $\Tilde{R}$ first increases w.r.t. small $N$ since the beamforming gain increases with a larger number of elements per RIS.
As $N$ grows large, $\Tilde{R}$ degrades because the distances between the UE and the RISs are statistically extended by the corresponding lower RIS density, which is equal to $\frac{\eta}{N}$.
The optimal $N^*$ in Fig.~\ref{Fig_ErgRate_Nopt} is obtained by a bisection method while that in Fig.~\ref{Fig_ErgRate_Nopt_typ} is calculated via the closed-form solution in \eqref{N_opt}.
The accuracy of both methods is verified by the numerical results.
Moreover, it can be observed that larger $C$ results in larger $N^*$, which is consistent with the observations in Remark~\ref{Re_OptN_1}.

Fig.~\ref{Fig_ErgRate_Nopt_rho1} shows the spatially ergodic rate versus $N$ under the constraint of $\lambda N \equiv 10$.
Different from Figs.~\ref{Fig_ErgRate_Nopt}--\ref{Fig_ErgRate_Nopt_typ}, random phase shifts are implemented at the RIS.
Similar to the observations from Figs.~\ref{Fig_ErgRate_Nopt}--\ref{Fig_ErgRate_Nopt_typ}, $\Tilde{R}$ increases w.r.t. small $N$ while decreases w.r.t. large $N$ for $\alpha_3=2$.
The optimal $N^*=45$ is calculated by \eqref{N_opt_2}, which is verified accurate to maximize $\Tilde{R}$.
On the other hand, $\Tilde{R}$ decreases monotonically w.r.t. $N$ for $\alpha_3=2.5,~3,$ and $3.5$. Consistent with \eqref{N_opt_3} in Lemma~\ref{lem_opt3}, the optimal $N^*$ equals 1.
It implies that the reflecting elements are suggested to spread out as much as possible under the condition of $\rho=1$ and large $\alpha_3$, without the consideration of deployment cost.

\begin{figure}[tb]
\centering\includegraphics[width=0.6\textwidth]{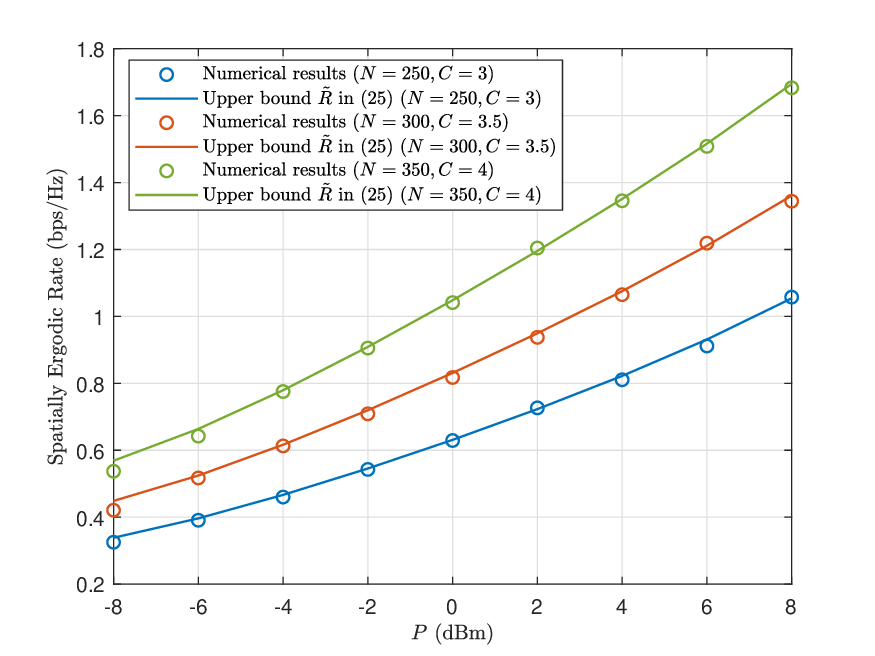}
\caption{Spatially ergodic rate versus $P$ with $\rho=0$ and $\lambda=0.005$.}
\label{Fig_ErgR_P}
\end{figure}

\begin{figure}[tb]
\centering\includegraphics[width=0.6\textwidth]{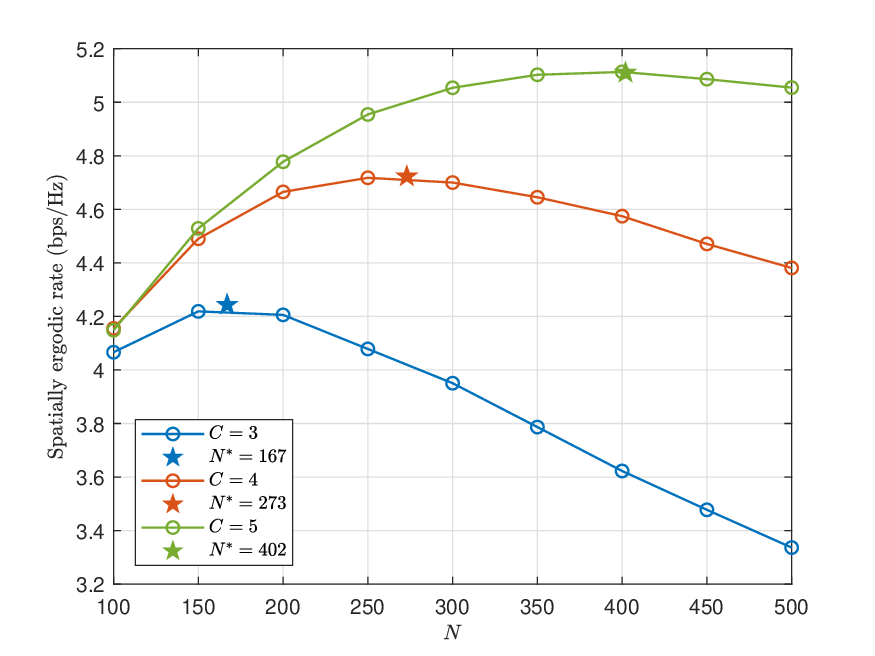}
\caption{Spatially ergodic rate under the constraint of $\eta=10$ with $\rho=0.1$ and $P=3$~dBm.}
\label{Fig_Nopt_lowSNR_rho01}
\end{figure}

\begin{figure}[tb]
\centering\includegraphics[width=0.6\textwidth]{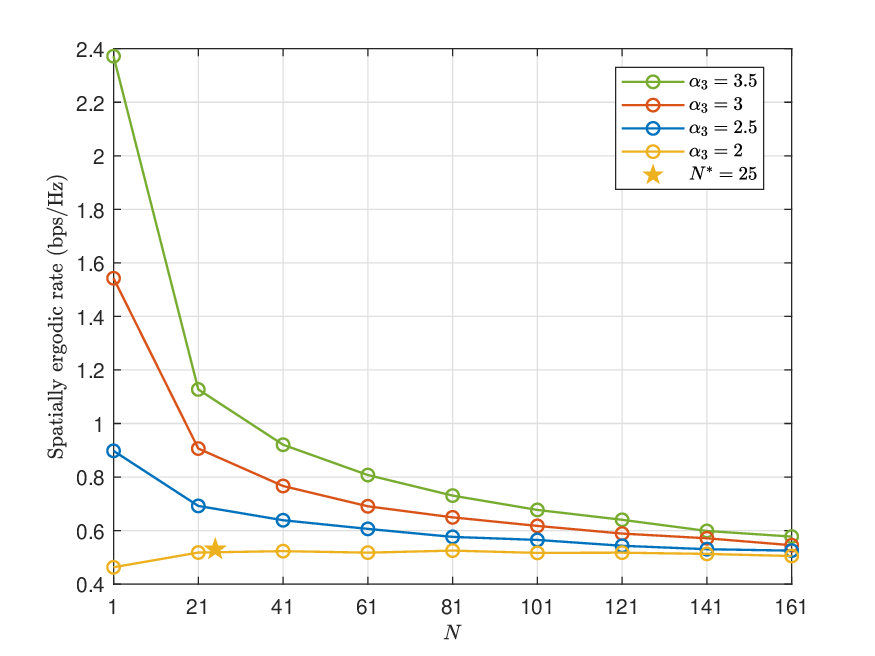}
\caption{Spatially ergodic rate with random phase shifts under the constraint of $\eta=10$ with $P=3$~dBm and $C=3$~m.}
\label{Fig_Nopt_lowSNR_rho1}
\end{figure}

In addition to the numerical results for high SNR in Figs.~\ref{Fig_ErgRate_C}$-$\ref{Fig_ErgRate_Nopt_rho1},
the spatially ergodic rate for low SNR is shown in Figs.~\ref{Fig_ErgR_P}-\ref{Fig_Nopt_lowSNR_rho1}.
In Fig.~\ref{Fig_ErgR_P}, lines correspond to the closed-form expression in \eqref{Eq_Erg_R_L} while markers correspond to Monte-Carlo simulations. The tightness of the derived ergodic rate is verified.
Moreover, Figs.~\ref{Fig_Nopt_lowSNR_rho01} and \ref{Fig_Nopt_lowSNR_rho1} display the spatially ergodic rate versus $N$ under the constraint of $\lambda N \equiv 10$ for fair comparison.
In the condition of bounded phase shift errors with $\rho=0.1$, it is observed from Fig.~\ref{Fig_Nopt_lowSNR_rho01} that $\Tilde{R}$ first increases as $N$ grows, due to the improved RIS beamforming gain, and then degrades as $N$ gradually increases, due to the corresponding lower RIS deployment density $\lambda$.
The optimal $N^*$ in Fig.~\ref{Fig_Nopt_lowSNR_rho01} is obtained by bisection search and the results are verified accurate.
On the other hand, random phase shifts are implemented at the RIS for the numerical results in Fig.~\ref{Fig_Nopt_lowSNR_rho1}.
Similar to the observations from Fig.~\ref{Fig_Nopt_lowSNR_rho01}, $\Tilde{R}$ increases w.r.t. small $N$ while decreases w.r.t. large $N$ for the particular case of $\alpha_3=2$.
However, for general cases of $\alpha_3=2.5,~3,$ and $3.5$, $\Tilde{R}$ decreases monotonically w.r.t. $N$, yielding $N^*=1$.
It implies that reduced RIS density is recommended in the case of $\rho=1$ and large $\alpha_3$, disregarding the deployment cost.
These observations provide engineering insights on the deployment of distributed RISs in practice.

\begin{figure}[tb]
\centering\includegraphics[width=0.6\textwidth]{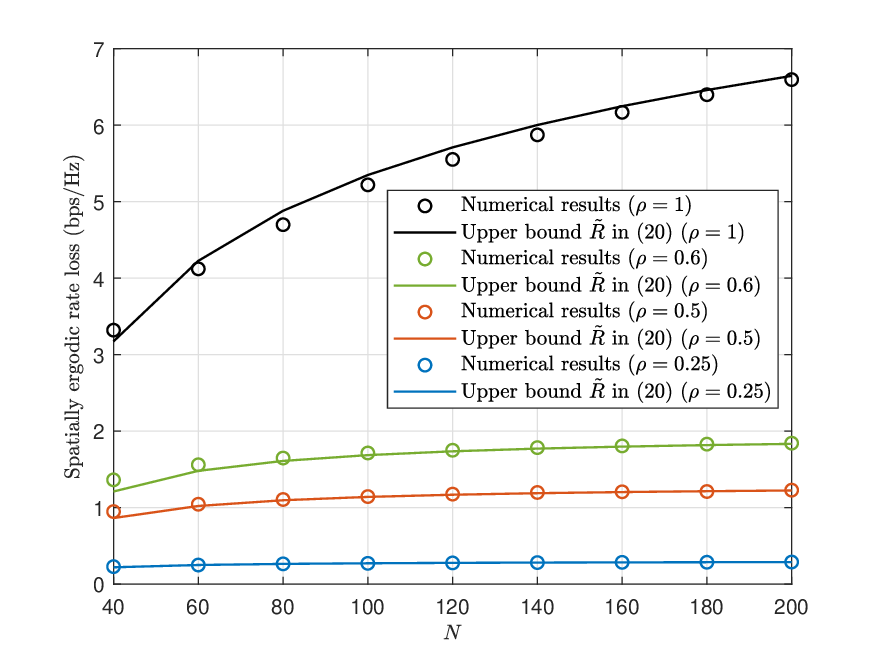}
\caption{Spatially ergodic rate loss versus $N$ with $P=15$~dBm, $\lambda=0.05$, and $C=10$~m.}
\label{Fig_ErgRateLoss_N}
\end{figure}

Fig.~\ref{Fig_ErgRateLoss_N} shows the spatially ergodic rate loss accounting for the phase shift errors compared to the ideal phase shifts.
From Fig.~\ref{Fig_ErgRateLoss_N}, the rate loss with $\rho=1$ logarithmically increases w.r.t. $N$ and goes to infinity as $N\rightarrow\infty$.
While for constrained errors with $\rho=0.25$, $0.5$, and $0.6$, the rate loss tends saturated at $0.3$, $1.2$, and $1.8$~bps/Hz for $N>40$, $120$, and $160$, respectively.
The upper bound of the rate loss in \eqref{Eq_ErgR_Loss_1} is also verified accurate.
It implies that the rate loss due to imperfect phase shifts is limited even for extremely large $N$, except for the random phase shifts.

\section{Conclusion}
\label{Sec_Con}
We derive a closed-form expression of the spatially ergodic rate for a multi-RIS-aided network. 
Under the deployment constraint of a total number of RIS elements per unit area, both the deployment density and the sizes of RISs are optimized for the ergodic rate maximization.
According to the derived closed-form solutions, it is suggested to form larger RISs with lower deployment density to support larger coverage and combat severer phase shift errors, under the condition of $\rho\in[0,1)$.
While for random phase shifts with $\rho=1$, the reflecting elements are suggested to spread out without the consideration of deployment cost.
Moreover, we analyze the ergodic rate loss due to the imperfect phase shifts in practice.
For bounded phase shift errors with $\rho\in[0,1)$, the rate loss tends to eventually reach a constant upper bound of $\left(1-\mathrm{e}^{-\pi\lambda C^2}\right)\log \frac{\pi^2\rho^2}{\sin^2(\rho\pi)}$ for extremely large RISs.
Note that for more advanced networks using other frequency reuse strategies, the effects of inter-cell and intra-cell interferences are interesting extensions as our future work.

\appendices
\section{Lemma~\ref{lemma_cos}}
\label{cosproof}

\begin{Lemma}
\label{lemma_cos}
For two independent phase errors $\tau_{n}\sim\mathcal{U}(-\rho\pi,\rho\pi)$ and $\tau_{m}\sim\mathcal{U}(-\rho\pi,\rho\pi)$, it follows that
\begin{align}
\mathbb{E}\left\{\cos\left(\nu_{n,m}\right)\right\}=\dfrac{\sin^2\left({\pi}\rho\right)}{{\pi}^2\rho^2},
\label{cos_nu}
\end{align}
where $\nu_{n,m}\triangleq\tau_{n}-\tau_{m}$ for $n\neq m$, $\forall n,m=1,...,N$.
\end{Lemma}

\begin{IEEEproof}
To start, we first derive the PDF of an intermediate random variable $Z=X-Y$ where $X\sim\mathcal{U}(-\rho\pi, \rho\pi)$ and $Y\sim\mathcal{U}(-\rho\pi, \rho\pi)$ are two
independent uniformly distributed variables.
When $-2\rho\pi<Z<0$, the cumulative distribution function (CDF) of $Z$ equals
\begin{align}
	F_{1}(Z)&=P(X<Z+Y)\notag\\
	&=\int_{-\rho\pi-Z}^{\rho\pi}\int_{-\rho\pi}^{Z+Y}\frac{1}{2\rho\pi}\frac{1}{2\rho\pi}\text{d}X\text{d}Y\notag\\
	&=\frac{1}{8a^2\pi^2}Z^2+\frac{1}{2\rho\pi}Z+\frac{1}{4},
\end{align}
which yields the PDF of $Z$ for $-2\rho\pi<Z<0$ as
\begin{align}
	f_{1}(Z)=\frac{\partial F_{1}(Z)}{Z}=\frac{1}{4\rho^2\pi^2}Z+\frac{1}{2\rho\pi}.
\end{align}
Analogously, we can obtain the PDF of $Z$ for $2\rho\pi>Z>0$ and get
\begin{numcases}
    {f(Z)=} {}
    \frac{1}{4\rho^2\pi^2}Z+\frac{1}{2\rho\pi},~~~~~-2\rho\pi<Z<0,
    \label{f1}\\
    -\frac{1}{4\rho^2\pi^2}Z+\frac{1}{2\rho\pi},~~~~2\rho\pi>Z>0.
    \label{f2}
\end{numcases}
Since the distribution of $\nu_{n,m}$ follow the PDF in \eqref{f1} and \eqref{f2}, we have
\begin{align}
	\mathbb{E}\left\{\cos\left(\nu_{n,m}\right)\right\}
	=&\int_{-2\rho\pi}^{0}\cos \nu \left( \frac{1}{4\rho^2\pi^2}\nu+\frac{1}{2\rho\pi}\right)\text{d} \nu
	  +\int_{0}^{2\rho\pi}\cos\nu \left( -\frac{1}{4\rho^2\pi^2}\nu+\frac{1}{2\rho\pi}\right)\text{d} \nu
	\notag\\
	=&-\frac{1}{2\rho^2\pi^2}\left(\nu\sin\nu+\cos\nu\right)\big|_{\nu=0}^{\nu=2\rho\pi}
 +\frac{1}{\rho\pi}\sin\nu\big|_{\nu=0}^{\nu=2\rho\pi}
	\label{Ecosv}
	.
\end{align}
Now, the expectation of $\mathbb{E}\left\{\cos\left(\nu_{n,m}\right)\right\}$ in \eqref{cos_nu} is directly obtained from \eqref{Ecosv}.
\end{IEEEproof}

\section{Proof of Theorem~\ref{Th_R_up}}
\label{Proof_Th1}
For notational simplicity, we define $z\triangleq\sum_{n=1}^N z_{n}\mathrm{e}^{j\tau_{n}}$ where $z_{n}\triangleq|g_{n}||h_{n }|$.
By using these definitions and applying the Jensen's inequality to \eqref{Eq_R1}, an upper bound of $R_1$ is obtained as
\begin{align}
    \bar{R}_1\triangleq &
    \log\left(1+ \frac{P}{\sigma^2}\left(\beta_l \beta_r \mathbb{E}\left\{|z|^2\right\}+\beta_d \mathbb{E}\left\{|h|^2\right\}
    +2\beta_l^\frac{1}{2}\beta_r^\frac{1}{2}\beta_d^\frac{1}{2}\mathbb{E}\left\{|h|\right\}\mathbb{E}\left\{\mathrm{Re}\{z\}\right\}\right)\right)
    \label{Eq_R1_up_1}
    .
\end{align}
In the following, we derive the terms $\mathbb{E}\left\{|h|\right\}$, $\mathbb{E}\left\{|h|^{2}\right\}$, $\mathbb{E}\left\{|z|^{2}\right\}$, and $\mathbb{E}\left\{\mathrm{Re}\{z\}\right\}$  in \eqref{Eq_R1_up_1} one-by-one by utilizing the statistical properties of the small-scale fading coefficients and the phase shift error.
Firstly, considering that $|h|$ follows the Rayleigh distribution $\mathcal{R}(\sqrt{1/2})$, we have $\mathbb{E}\left\{|h|\right\} = \frac{\sqrt{\pi}}{2}$ and $\mathbb{E}\left\{|h|^2\right\}=1$.

Next, according to the definition $z\triangleq\sum_{n=1}^N z_{n}\mathrm{e}^{j\tau_{n}}$, $\mathbb{E}\left\{|z|^{2}\right\}$ is calculated as
\begin{align}
	\mathbb{E}\left\{|z|^2\right\}
	=&\mathbb{E}\!\left\{\!\left( \sum_{n=1}^N\! z_{n}\mathrm{e}^{j\tau_{n}}\right)
	\left( \sum_{n=1}^N\!\! z_{n}\mathrm{e}^{j\tau_{n}}\!\right)^{*}\right\}
	\nonumber\\
	=&\mathbb{E}\!\left\{ \sum_{n=1}^N z_{n}^2
	+\sum_{n=1}^N \sum_{m\neq n} z_{n}z_{m}\cos\left(\nu_{n,m}\right)\!\right\}
	\label{zzall}\\
	=& \sum_{n=1}^N	 \mathbb{E}\left\{z_{n}^2\right\}+	\sum_{n=1}^N \sum_{m\neq n} \mathbb{E}\!\left\{\!z_{n}\!\right\}\!\mathbb{E}\left\{\!z_m\!\right\}\!\mathbb{E}\!\left\{\!\cos\left(\nu_{n,m}\right)\right\}
	\label{HH}\\
	=& N	+\frac{\sin^2(\rho\pi)}{16\rho^2} N(N-1),
 	\label{z_secondterm}
\end{align}
where
\eqref{zzall} utilizes the Euler's formula, i.e.,
$\mathrm{e}^{j\tau_{n}}\mathrm{e}^{-j\tau_m}+\mathrm{e}^{j\tau_m}\mathrm{e}^{-j\tau_{n}}=\cos(\nu_{n,m})+\cos(\nu_{m,n})$ where $\nu_{n,m}\triangleq\tau_{n}-\tau_m$ is defined as the difference between two phase errors,
\eqref{HH} comes from the fact that $z_{n}$, $z_{m}$, and $\nu_{n,m}$ are independent with each other for $ m\neq n$,
and \eqref{z_secondterm} uses $\mathbb{E}\left\{\cos\left(\nu_{n,m}\right)\right\}=\dfrac{\sin^2\left({\pi}\rho\right)}{{\pi}^2\rho^2}$ in Lemma~\ref{lemma_cos} and the statistical properties of a double-Rayleigh distributed random variable $z_{n}$, i.e.,
$\mathbb{E}\{z_{n}\}\triangleq\frac{\pi}{4}$ and
$\mathbb{V}\{z_{n}\}\triangleq 1-\frac{\pi^2}{16}$,
which are obtained from \cite[Eqs. (10), (11)]{jlyu}.

Similarly, the term $\mathbb{E}\left\{\mathrm{Re}\{z\}\right\}$ is evaluated as
\begin{align}
    \mathbb{E}\left\{\mathrm{Re}\{z\}\right\} = &
    \mathbb{E}\left\{ \sum_{n=1}^N z_{n} \cos\left(\tau_{n}\right)\right\}
    \nonumber\\=&
    \sum_{n=1}^N \mathbb{E}\left\{z_{n}\right\} \mathbb{E}\left\{\cos\left(\tau_{n}\right)\right\}
    \label{Eq_zRe_Ind}\\=&
    \frac{\sin\left(\rho\pi\right)}{4\rho}N,
    \label{Eq_zRe}
\end{align}
where \eqref{Eq_zRe_Ind} comes from the fact that $z_n$ and $\tau_n$ is independent of each other and \eqref{Eq_zRe} uses $\mathbb{E}\{z_{n}\}\triangleq\frac{\pi}{4}$ and $\mathbb{E}\left\{\cos\left(\tau_{n}\right)\right\}=\frac{\sin\left(\rho\pi\right)}{\rho\pi}$, which is directly obtained from $\tau_{n}\sim\mathcal{U}(-\rho\pi,\rho\pi)$.

Now that, the upper bound in \eqref{Eq_R1_up} is obtained by substituting $\mathbb{E}\left\{|h|\right\} = \frac{\sqrt{\pi}}{2}$, $\mathbb{E}\left\{|h|^2\right\}=1$, \eqref{z_secondterm}, and \eqref{Eq_zRe} into \eqref{Eq_R1_up_1} and using the definition of $\mu$ in \eqref{Eq_mu}.
Analogously, the upper bound in \eqref{Eq_R2_up} is obtained by applying the Jensen's inequality to \eqref{Eq_R2} and using $\mathbb{E}\left\{|h|^2\right\}=1$.

\section{Preliminary Results for the Proof of Theorems~\ref{Th_Erg_R_int}$-$\ref{Th_Erg_R_lowSNR}}
\label{Ap_Useful}
In this appendix, we provide the derivations of the mathematical expectations w.r.t $d$ and $r$, which are useful for the proof of Theorems~\ref{Th_Erg_R_int}$-$\ref{Th_Erg_R_lowSNR}.
Before calculating the definite integrals over the corresponding probability distributions, we first give the PDFs of $r$ and $d$.
From the null probability of an HPPP with density $\lambda$, the probability of $r>x$ is given as
\begin{align}
    \text{P}\{r>x\} = \mathrm{e}^{-\pi\lambda x^2},~~~x\geq 0,
\end{align}
which gives the CDF of $r$ as
\begin{align}
    F_r\{x\} = 1-\text{P}\{r>x\} = 1- \mathrm{e}^{-\pi\lambda x^2},~~~x\geq 0.
\end{align}
Then, the PDF of $r$ is directly derived as
\begin{align}
    f_r(x) = \frac{\partial F_r\{x\}}{\partial x} = 2\pi\lambda x \mathrm{e}^{-\pi\lambda x^2},~~~x\geq 0.
    \label{PDF_r}
\end{align}
On ther other hand, assuming that the UE is uniformly located within the edge of a cell, the PDF of $d$ is
\begin{align}
     f_d(d)= \frac{2d}{D_2^2-D_1^2},~~~D_1\leq d \leq D_2.
     \label{PDF_d}
\end{align}

Using \eqref{PDF_d}, the expectation of $\log d$  is evaluated as
\begin{align}
    \mathbb{E}_d\left\{\log d\right\}
    &= \frac{1}{\ln 2} \int_{D_1}^{D_2} \frac{2x}{D_2^2-D_1^2} \ln x  ~\mathrm{d}x
    \nonumber\\&=
    \frac{1}{\ln 2}\frac{1}{D_2^2-D_1^2} \int_{D_1}^{D2}  \ln x  ~\mathrm{d}x^2
    \nonumber\\&=
    \frac{1}{\ln 2}\frac{1}{D_2^2-D_1^2} \left( x^2\ln x \big|_{D_1}^{D2} - \int_{D_1}^{D2} x^2 ~\mathrm{d} \ln x\right)
    \nonumber\\&=
    \frac{1}{\ln 2} \left(\frac{D_2^2\ln D_2 -D_1^2\ln D_1 }{D_2^2-D_1^2} -\frac{1}{2}\right)
    \label{Eq_Int_d}
    .
\end{align}
Similarly, using \eqref{PDF_r}, we derive the expectation of $\log r$ for $r\in[0,C]$ as
\begin{align}
    \mathbb{E}_r\left\{\log r|r\in[0,C]\right\}
    =& \frac{1}{\ln 2} \int_{0}^{C} 2\pi\lambda r  \mathrm{e}^{-\pi\lambda r^2}\ln r ~\mathrm{d}r
    \nonumber\\=&
    \frac{\pi\lambda}{2\ln 2} \int_{0}^{C^2}  \mathrm{e}^{-\pi\lambda x} \ln x ~\mathrm{d}x
    \label{Eq_Int_ln_r_1}\\=&
    \frac{1}{2\ln 2} \left[ \mathrm{Ei} \left(-\pi\lambda x\right)\big|_{0}^{C^2} -\mathrm{e}^{-\pi\lambda x} \ln x\big|_{0}^{C^2} \right]
    \label{Eq_Int_ln_r_2}\\=&
    \frac{1}{2\ln 2}
    \left[\mathrm{Ei} \left(-\pi\lambda C^2\right)
    \!-\!\mathrm{e}^{-\pi\lambda C^2} \ln C^2
    \!-\!\ln(\pi\lambda)\!-\!E_0\right]
    \label{Eq_Int_ln_r}
    ,
\end{align}
where \eqref{Eq_Int_ln_r_1} uses $x= r^2$, \eqref{Eq_Int_ln_r_2} comes from \cite[Eq. (2.751)]{table},
and \eqref{Eq_Int_ln_r} comes from $\ln( 0_+)-\mathrm{Ei} \left(-\pi\lambda 0_+\right) = \ln( 0_+)- \ln(\pi\lambda 0_+) -E_0= -\ln(\pi\lambda) -E_0$ in \cite[Eq. (8.214)]{table}.
In addition, the probabilities of $r\in[0,C]$ and $r\in(C,\infty]$ are respectively characterized as
\begin{align}
    \int_{0}^{C}f_r(r)~\mathrm{d}r
    = \int_{0}^{C} 2\pi\lambda r \mathrm{e}^{-\pi\lambda r^2} ~\mathrm{d}r
    = 1-\mathrm{e}^{-\pi\lambda C^2},
    \label{Eq_Int_r_1}
\end{align}
and
\begin{align}
    \int_{C}^{\infty}f_r(r)~\mathrm{d}r
    = \int_{C}^{\infty} 2\pi\lambda r \mathrm{e}^{-\pi\lambda r^2} ~\mathrm{d}r
    = \mathrm{e}^{-\pi\lambda C^2}.
    \label{Eq_Int_r_2}
\end{align}

\section{Proof of Theorem~\ref{Th_Erg_R_int}}
\label{Proof_Th_R_int}
To begin with, we first consider the case of $r\in[0,C]$, where the UE is served by both the RIS and BS.
According to \eqref{Eq_beta} and \eqref{Eq_R1_up}, the spatially ergodic rate is derived as
\begin{align}
    \Tilde{R}_{1}&=
	\mathbb{E}_{l,d,r}\!\!\left\{\!\log\!\left( \!\!1\!\!+\!\frac{P}{\sigma^2}\beta\!\left[\beta l^{-\alpha_2} r^{-\alpha_3} \mu^2 N^2\!\!+\!\!\beta l^{-\alpha_2} r^{-\alpha_3}\left(\!1\!-\!\mu^2\right)\!N
	\!\!+\!\sqrt{\!	\beta l^{-\alpha_2} r^{-\alpha_3}d^{-\alpha_1} \pi}\mu N\!+\! d^{-\alpha_1}\!\right]\!\!\right)\!\!\right\}
	\nonumber\\ &=
	\mathbb{E}_{d,r}\left\{\!\log\left(1+\frac{P}{\sigma^2}\beta d^{-\alpha_2} r^{-\alpha_3}\left[\beta \mu^2 N^2+\beta\left(1\!-\!\mu^2\right)N
	\!+\!\sqrt{	\beta r^{\alpha_3}d^{\alpha_2-\alpha_1} \pi}\mu N
	\!+\! d^{\alpha_2-\alpha_1} r^{\alpha_3}\right]\!\right)\!\!\right\}
	\label{Eq_R1_Er_1_lowSNR}\\ &=
	\log\left[\frac{P}{\sigma^2}\beta^2N\left(\mu^2 N+1-\mu^2\right) \right]\int_{0}^{C}f_r(r)~\mathrm{d}r
	-\alpha_2\int_{D_1}^{D_2}f_d(d)\log d~\!\mathrm{d}d \int_{0}^{C}f_r(r)~\mathrm{d}r
	\nonumber\\&~~~~~~~
        -\alpha_3\int_{0}^{C}f_r(r)\log r~\mathrm{d}r
	+\mathcal{G}(N,\mu)
	\label{Eq_R1_Er_4_lowSNR}\\&=
	\left(1-\mathrm{e}^{-\pi\lambda C^2}\right)
	\left[\log\left(\frac{P}{\sigma^2}\beta^2N\left(\mu^2 N+1\!-\!\mu^2\right) \right)
	-\frac{\alpha_2}{\ln 2} \left(\frac{D_2^2\ln D_2 -D_1^2\ln D_1 }{D_2^2-D_1^2} -\frac{1}{2}\right)\right]
	\nonumber\\&~~~~~~~
	-\frac{\alpha_3}{2\ln 2}
    \left[\mathrm{Ei} \left(-\pi\lambda C^2\right)
    -\mathrm{e}^{-\pi\lambda C^2} \ln C^2
    -\ln(\pi\lambda)-E_0\right]
    +\mathcal{G}(N,\mu)
    ,\label{Eq_R1_Er_int}
\end{align}
where \eqref{Eq_R1_Er_1_lowSNR} uses the relationship $l=d$ for the cell-edge UE, \eqref{Eq_R1_Er_4_lowSNR} uses $\int_{D_1}^{D_2} f_d(d) ~\mathrm{d}d = 1$ and the definition of $\mathcal{G}(N,\mu)$ in \eqref{Eq_G_int} along with the PDFs in \eqref{PDF_r} and \eqref{PDF_d},
and \eqref{Eq_R1_Er_int} utilizes the derived preliminary results \eqref{Eq_Int_d}, \eqref{Eq_Int_ln_r}, and \eqref{Eq_Int_r_1} in Appendix~\ref{Ap_Useful}.

Then, we consider the case of $r\in(C,\infty)$, where the UE is solely associated to the BS.
By substituting \eqref{Eq_beta} and \eqref{Eq_R2_up} into \eqref{Eq_Er_Exp}, the spatially ergodic rate is expressed as
\begin{align}
    \Tilde{R}_2 &=
    \mathbb{E}_{d,r}\left\{\log\left(1+ \frac{P}{\sigma^2} \beta d^{-\alpha_1} \right)\right\}
    \nonumber\\&
    =\int_{D_1}^{D_2} \!\!\!\!\int_{C}^{\infty}\!\!\!\log\left(\!\!1 \!+ \!\frac{P}{\sigma^2} \beta d^{-\alpha_1}\! \!\right)   f_r(r)  f_d(d)~\mathrm{d}r\mathrm{d}d
    .
 \label{Eq_R2_Er}
\end{align}
Then, using the PDFs in \eqref{PDF_r} and \eqref{PDF_d}, the expression of $\Tilde{R}_2$ is obtained in \eqref{Eq_R2_Er_int}.
Finally, the spatially ergodic rate in \eqref{Eq_Erg_R_int} is obtained by substituting \eqref{Eq_R1_Er_int} and \eqref{Eq_R2_Er_int} into $\Tilde{R} = \Tilde{R}_1+\Tilde{R}_2$.

\section{Proof of Theorem~\ref{Th_Erg_R}}
\label{Proof_Th2}

For high SNR, we use the result in Theorem~\ref{Th_Erg_R_int} and first consider the case with $r\in[0,C]$.
Under this condition, the spatially ergodic rate is given in \eqref{Eq_R1_Er_int} and we need to further derive the closed-form expression of $\mathcal{G}(N,\mu)$.
After removing the negligible term in \eqref{Eq_G_int} due to the large value of $\frac{P}{\sigma^2}$, $\mathcal{G}(N,\mu)$ is rewritten as
\begin{align}
    \mathcal{G}_H(N,\mu) & =
	\int_{D_1}^{D_2}\int_{0}^{C}\log\left(1
	+ \frac{\sqrt{\pi\beta d^{\alpha_2-\alpha_1}r^{\alpha_3} }\mu N
	+ d^{\alpha_2-\alpha_1} r^{\alpha_3}}
	{\beta\mu^2 N^2+\beta\left(1-\mu^2\right)N} \right)2\pi\lambda r  \mathrm{e}^{-\pi\lambda r^2} \frac{2d}{D_2^2-D_1^2}~\mathrm{d}r\mathrm{d}d
	\label{Eq_G_highSNR}\\& <
	\frac{1}{\ln 2}\int_{D_1}^{D_2}\int_{0}^{C}
	\frac{\sqrt{\pi\beta d^{\alpha_2-\alpha_1}r^{\alpha_3} }\mu N
	+ d^{\alpha_2-\alpha_1} r^{\alpha_3}}{\beta\mu^2 N^2+\beta\left(1-\mu^2\right)N}
	2\pi\lambda r  \mathrm{e}^{-\pi\lambda r^2}\frac{2d}{D_2^2-D_1^2}~\mathrm{d}r\mathrm{d}d
	\label{Eq_G_5}\\&=
    \frac{1}{\beta\ln 2}
	\left(\frac{ \kappa_1 \sqrt{\pi\beta}\mu\gamma\left(\frac{\alpha_3}{4}+1,\pi\lambda C^2\right) }{(\pi\lambda)^{\frac{\alpha_3}{4}} (\mu^2 N +1-\mu^2)}
	+\frac{\kappa_2\gamma\left(\frac{\alpha_3}{2}+1,\pi\lambda C^2\right)}{(\pi\lambda)^{\frac{\alpha_3}{2}}N \left(\mu^2 N +1-\mu^2\right)}  \right)
	.\label{Eq_G_6}
\end{align}
Since the closed-form expression of $\mathcal{G}_H(N,\mu)$ is still hard to acquire, we resort to a tight upper bound for moderate-to-large $N$ expressed in \eqref{Eq_G_6},
where \eqref{Eq_G_5} uses $\ln(1+x) < x$ for $x>0$,
and \eqref{Eq_G_6} utilizes \cite[Eq. (3.381.8)]{table} and the definitions in \eqref{ka1} and \eqref{ka2}.
Now that, by substituting \eqref{Eq_G_6} into \eqref{Eq_R1_Er_int}, the desired expression of $\Tilde{R}_{1}$ for high SNR is obtained.

Then, we consider the condition with $r\in(C,\infty)$.
According to the spatially ergodic rate $\Tilde{R}_2$ in \eqref{Eq_R2_Er}, the closed-form rate expression at high SNR is derived as
\begin{align}
    \Tilde{R}_2 = \!&
    \int_{D_1}^{D_2} \!\!\!\!\int_{C}^{\infty}\!\left[\log\left(\frac{P}{\sigma^2}\beta \right)-\alpha_1\log d\right]  f_r(r)  f_d(d)~\mathrm{d}r\mathrm{d}d
    \label{Eq_R2_Er_2}\\=&
    \log\left(\frac{P}{\sigma^2}\beta \right)\int_{C}^{\infty}f_r(r)~\mathrm{d}r
	-\alpha_1\int_{D_1}^{D_2}f_d(d)\log d~\mathrm{d}d \int_{C}^{\infty}f_r(r)~\mathrm{d}r
	\label{Eq_R2_Er_3}\\=&
	\mathrm{e}^{-\pi\lambda C^2}
	\left[\log\left(\!\frac{P}{\sigma^2}\beta \!\right)\!-\! \frac{\alpha_1}{\ln 2} \!\!\left(\!\!\frac{D_2^2\ln D_2 \!-\!D_1^2\ln D_1 }{D_2^2-D_1^2} \!-\!\frac{1}{2}\!\right)\!\right]
	\label{Eq_R2_Er_4}\!\!,
\end{align}
where
\eqref{Eq_R2_Er_3} comes from that $\int_{D_1}^{D_2} f_d(d) ~\mathrm{d}d = 1$
and \eqref{Eq_R2_Er_4} uses \eqref{Eq_Int_d} and \eqref{Eq_Int_r_2}.
Finally, the spatially ergodic rate in \eqref{Eq_Erg_R} is obtained by substituting \eqref{Eq_R1_Er_int}, \eqref{Eq_G_6}, and \eqref{Eq_R2_Er_4} into $\Tilde{R} = \Tilde{R}_1+\Tilde{R}_2$.

\section{Proof of Theorem~\ref{Th_Erg_R_lowSNR}}
\label{Proof_Th3}

For low SNR, we use the result in Theorem~\ref{Th_Erg_R_int} and first consider the case of $r\in[0,C]$.
Similarly to the case of high SNR, the spatially ergodic rate is expressed in \eqref{Eq_R1_Er_int} and we focus on the term $\mathcal{G}(N,\mu)$ involving integrals.
Compared to $\mathcal{G}_H(N,\mu)$ in \eqref{Eq_G_highSNR} for high SNR, a small value of $\frac{P}{\sigma^2}$ introduces an additional term that is no longer negligible in $\mathcal{G}_L(N,\mu)$, as expressed below.
\begin{align}
    &\mathcal{G}_L(N,\mu)
    \nonumber\\
    =&
    \int_{D_1}^{D_2}\!\!\!\!\!\int_{0}^{C}\!\log\!\!\left(\!\!1
	\!\!+ \!\!\frac{\sqrt{\pi\beta d^{\alpha_2\!-\!\alpha_1}r^{\alpha_3} }\mu N
	\!+\! d^{\alpha_2\!-\!\alpha_1} r^{\alpha_3} + (\sigma^2/P/\beta) d^{\alpha_2} r^{\alpha_3} }
	{\beta\mu^2 N^2+\beta\left(1-\mu^2\right)N} \!\!\right)\!2\pi\lambda r  \mathrm{e}^{-\pi\lambda r^2} \frac{2d}{D_2^2-D_1^2}~\mathrm{d}r\mathrm{d}d
     \label{Eq_G_L}
 \\<&
    \frac{1}{\beta\ln 2}\!\!
	\left(\!\frac{ \kappa_1 \sqrt{\pi\beta}\mu\gamma\left(\frac{\alpha_3}{4}\!+\!1,\pi\lambda C^2\right) }{(\pi\lambda)^{\frac{\alpha_3}{4}} (\mu^2 N \!+\!1-\mu^2)}
	\!+\!\frac{\kappa_2\gamma\left(\frac{\alpha_3}{2}+1,\pi\lambda C^2\right)}{(\!\pi\lambda\!)^{\frac{\alpha_3}{2}}N \left(\mu^2 N \!+\!1\!\!-\!\mu^2\!\right)}
       \! +\!\frac{\kappa_3\gamma\left(\frac{\alpha_3}{2}+1,\pi\lambda C^2\right)}{(\!P\beta/\sigma^2)(\pi\lambda)^{\frac{\alpha_3}{2}}\!N \!\left(\mu^2 N\! +\!1\!\!-\!\mu^2\right)}\!\right)
	\!\!.\label{Eq_G_L_6}
\end{align}
Since it is intractable to get a closed-form expression for $\mathcal{G}_L(N,\mu)$, we resort to an upper bound derived in \eqref{Eq_G_L_6}, which is obtained by the same mathematical derivations in \eqref{Eq_G_6} and using the definitions in \eqref{ka1}, \eqref{ka2}, and \eqref{ka3}.
Now that, by substituting \eqref{Eq_G_L_6} into \eqref{Eq_R1_Er_int}, the desired expression of $\Tilde{R}_{1}$ for low SNR is obtained.

Then, we consider the case of $r\in(C,\infty)$.
By substituting the PDFs in \eqref{PDF_r} and \eqref{PDF_d} into \eqref{Eq_R2_Er}, the spatially ergodic rate $\Tilde{R}_2$ at low SNR is
\begin{align}
    \Tilde{R}_2 &= \!
    \int_{D_1}^{D_2} \!\!\!\!\int_{C}^{\infty}\!\!\!\log\left(\!\!1 \!+ \!\frac{P}{\sigma^2} \beta d^{-\alpha_1}\! \!\right)  2\pi\lambda r  \mathrm{e}^{-\pi\lambda r^2} \frac{2d}{D_2^2\!-\!D_1^2}~\mathrm{d}r\mathrm{d}d
    \nonumber\\&
    < \frac{1}{\ln 2} \int_{D_1}^{D_2} \!\!\!\!\int_{C}^{\infty} \left(\frac{P}{\sigma^2} \beta d^{-\alpha_1} \!\!\right)  2\pi\lambda r  \mathrm{e}^{-\pi\lambda r^2} \frac{2d}{D_2^2\!-\!D_1^2}~\mathrm{d}r\mathrm{d}d
    \label{Eq_R2_Er_3_lowSNR}\\&
    =  \frac{P\beta}{\sigma^2\ln 2 }  \mathrm{e}^{-\pi\lambda C^2} \frac{2\left(D_2^{2-\alpha_1}-D_1^{2-\alpha_1}\right)}{(2-\alpha_1)\left(D_2^2-D_1^2\right)}
	\label{Eq_R2_Er_lowSNR},
\end{align}
where
\eqref{Eq_R2_Er_3_lowSNR} utilizes the fact that $\ln(1+x) < x$ for $x>0$.
Finally, the spatially ergodic rate in \eqref{Eq_Erg_R_L} is obtained by substituting \eqref{Eq_R1_Er_int}, \eqref{Eq_G_L_6}, and \eqref{Eq_R2_Er_lowSNR} into $\Tilde{R} = \Tilde{R}_1+\Tilde{R}_2$.

\ifCLASSOPTIONcaptionsoff
  \newpage
\fi

\end{document}